\documentclass{article}

\PassOptionsToPackage{numbers, compress}{natbib}
\usepackage[final]{neurips_2021}
\usepackage{caption}
\usepackage{subcaption}
\usepackage[utf8]{inputenc} % allow utf-8 input
\usepackage[T1]{fontenc}    % use 8-bit T1 fonts
\usepackage{hyperref}       % hyperlinks
\usepackage{url}            % simple URL typesetting
\usepackage{booktabs}       % professional-quality tables
\usepackage{amsfonts}       % blackboard math symbols
\usepackage{nicefrac}       % compact symbols for 1/2, etc.
\usepackage{microtype}      % microtypography
\usepackage{xcolor}         % colors
\usepackage{graphicx}
\usepackage{wrapfig}
\usepackage{gensymb}
\usepackage{wrapfig}
\usepackage{pdfpages}
\usepackage{float}

\usepackage{bm}
\usepackage{booktabs}
\usepackage{siunitx}
\usepackage{mathpazo}
\usepackage{multirow}
\newcommand{\rpm}{\raisebox{.2ex}{$\scriptstyle\pm$}}

\title {EEGEyeNet: a Simultaneous Electroencephalography and Eye-tracking Dataset and Benchmark for Eye Movement Prediction
}
\author{Ard Kastrati$^{1}$\thanks{Both authors contributed equally} $\;$ \textbf{Martyna Beata Płomecka$^{2}$\footnotemark[1]} $\;$ \textbf{Dami\'an Pascual}$^{1}$ $\;$ \textbf{Lukas Wolf}$^{1}$ $\;$ \\ \textbf{Victor Gillioz}$^{1}$ $\;$ \textbf{Roger Wattenhofer}$^{1}$ $\;$ \textbf{Nicolas Langer}$^{2}$ \\
 $^{1}$ETH Zurich, Switzerland \\
 $^{2}$University of Zurich, Switzerland \\
 \texttt{akastrati@ethz.ch, martyna.plomecka@uzh.ch}
 }

\begin{document}

\maketitle

\begin{abstract}
  %The abstract paragraph should be indented \nicefrac{1}{2}~inch (3~picas) on
  %both the left- and right-hand margins. Use 10~point type, with a vertical
  %spacing (leading) of 11~points.  The word \textbf{Abstract} must be centered,
  %bold, and in point size 12. Two line spaces precede the abstract. The abstract
  %must be limited to one paragraph.
 We present a new dataset and benchmark with the goal of advancing research in the intersection of brain activities and eye movements. Our dataset, EEGEyeNet, consists of simultaneous Electroencephalography (EEG) and Eye-tracking (ET) recordings from 356 different subjects collected from three different experimental paradigms. Using this dataset, we also propose a benchmark to evaluate gaze prediction from EEG measurements. The benchmark consists of three tasks with an increasing level of difficulty: left-right, angle-amplitude and absolute position. We run extensive experiments on this benchmark in order to provide solid baselines, both based on classical machine learning models and on large neural networks. We release our complete code and data and provide a simple and easy-to-use interface to evaluate new methods. 
\end{abstract}

\section{Introduction}

%1. Show importance of eye tracking (independent of eeg)
Tracking eye position is a subject of active research due to its multiple applications across different fields such as behavioural science \cite{eckstein2017beyond}, assistive technology \cite{ken}, or user experience \cite{bergstrom2014eye}. Eye tracking in combination with conventional research methods, like behavioral measures, can help assess and potentially diagnose neurological diseases such as Autism Spectrum Disorder \cite{falck2013eye}, Obsessive Compulsive Disorder \cite{bradley2016obsessive}, Schizophrenia \cite{holzman1973eye}, Parkinson‘s \cite{shibasaki1979oculomotor}, and Alzheimer‘s disease \cite{hutton1984eye}. Additionally, eye tracking technology can be used to detect states of drowsiness \cite{devi2008driver}, to support communication for locked-in patients \cite{abu2006design}, and to measure attention in marketing~\cite{wedel2008eye}. 
%2. Show importance where combination of eeg and eye is important and thus importance of our dataset.
In the last decade, technological advances have allowed complementing eye-tracking technology with EEG --- a non-invasive, minimally restrictive, and,  relatively low-cost measure of mesoscale brain dynamics with high temporal resolution. Combining behavioral information gained from eye tracking with the neurophysiological markers provided by EEG enables researchers to study perceptual, attentional or cognitive processes in naturalistic situations \cite{konig2012combining}. %Thus, it is expected that there will be an ever growing demand for methods that estimate gaze position and aligning brain activation with current visual information processing. 
%Interestingly, our previous work indicates that it might be possible to infer gaze direction directly from EEG data \cite{10.1145/3448018.3458014}.
%recent work has shown that it is possible to infer gaze direction directly from EEG data \cite{10.1145/3448018.3458014}. 
Notably, estimating gaze position using EEG would 
%enable extracting such behavioral information with no extra costs for ET hardware and setup. This way, EEG-based eye tracking would 
make available gaze position information in a wide variety of studies that cannot acquire eye tracking data otherwise, either because of the unavailability of ET hardware or the lack of in-house expertise. This, in turn, could accelerate scientific discovery on human behavior and neurological and psychiatric diseases,
%perceptual, attentional and cognitive processes, 
particularly during free viewing of complex stimuli (i.e. naturalistic paradigms) and in clinical settings, in which the installation of an eye-tracker is impractical (e.g. hospital bed). 

Advancing cognitive research at the intersection of brain dynamics and eye movement requires synchronized data from EEG and eye-tracking.
Such data contains gaze pattern of eye movements recorded with eye- tracking as well as the neurophysiological markers provided by EEG, allowing researchers to study attention and reaction time 
%controlling fixation, measuring saccadic reaction times
\cite{plochl2012combining}, or to improve brain-computer interfaces \cite{ma2018combining}.
In the same line, estimating gaze position from EEG signals is typically approached with machine learning and deep learning models \cite{shishkin2016eeg,10.1145/3448018.3458014}, which need significant amounts of annotated data for training. 
However, collecting and annotating simultaneous EEG and eye position data is time-consuming and expensive since it requires equipment and expertise for both EEG acquisition and eye-tracking. Hence, the access to concurrently recorded EEG-ET data is highly restricted, which significantly slows down progress in this field. %research in the intersection of EEG and eye movement. F
%urthermore, estimating gaze position from EEG signals is a challenging problem typically approached with machine learning and deep learning models \cite{shishkin2016eeg,10.1145/3448018.3458014}, which need significant amounts of accurately labelled data for training.
To help bridging this gap, we release EEGEyeNet, a large dataset of EEG data synchronized with precise eye-tracking recordings. % collected from 357 subjects following three different experimental paradigms. 
%Given that EEG-based gaze estimation is a particularly relevant application, 
%Furthermore, there are currently no reliable methods for estimating gaze position from EEG. 
Furthermore, given the multiple benefits that EEG-based eye tracking can bring to different domains, in conjunction with our dataset, we present a benchmark for evaluating gaze estimation from EEG. This benchmark comprises three tasks of increasing difficulty and is conceived as a tool to facilitate comparable and reproducible research on gaze estimation from EEG data.
%to aims to evaluate and compare gaze estimation methods from EEG data. 
%To facilitate comparable and reproducible experimentation, in conjunction with our dataset, we present a benchmark comprising three well-defined tasks that aim to evaluate and compare gaze estimation methods from EEG data. 
We run extensive experiments on the proposed benchmark to establish baseline performance. In order to foster further research in this field we make all of our code and infrastructure available in the following site: \url{http://www.eegeye.net}. 
To conclude, our key contributions can be summarized as:
%3. Bulletpoints that summarize it
\begin{itemize}
    \item A dataset of high-density 128-channel EEG data synchronized with video-infrared eye tracking from 356 healthy adults that amounts to a total of more than 47 hours of recording. %approximately 2842 minutes of recording.
    Along with the raw dataset we provide preprocessed data and code for preprocessing and feature extraction.
    %Along with the dataset we release different preprocessing and feature extraction methods.% to ease data preparation. 
    %performing three different experimental paradigms measuring eye movements and amounting to a total of XXXX minutes of recording along with a set of preprocessing and feature extraction methods.  
    %\item A set of preprocessing methods and already extracted features and events in both the EEG and Eye-Tracking recordings along with a general data preparation module to facilitate the use of this dataset for various research goals. 
    \item A benchmark for gaze estimation from EEG signals that consists of three evaluation tasks with increasing difficulty. This benchmark is built on a subset of the EEGEyeNet dataset.% and we provide the necessary data preparation and preprocessing modules. % to preprocess and prepare the dataset for several benchmarking tasks.
    %This benchmark constitutes a well-defined framework for comparing EEG-based eye tracking methods and ... data preparation module to prepare the dataset for several benchmarking tasks.
    \item An extensive experimentation that establishes baseline performance on the proposed benchmark.
    %providing clean preprocessed data (as well as raw samples) establishing the baseline performance on the proposed benchmark. 
\end{itemize}

\section{Related Work}
Gaze prediction is an active research topic with applications in human behaviour analysis \cite{chaaraoui2012review}, advertisement \cite{okada2018advertisement}, and human-computer interaction \cite{majaranta2014eye}, to name a few. 
%The recent development of gaze prediction builds on decades of research. 
Previous research found evidence suggesting that action selection is facilitated by attention \cite{eckstein2017beyond}. 
Furthermore, \cite{eth} demonstrated the possibility of performing activity recognition from eye movements. 
To predict gaze location, some models use saliency maps \cite{nakashima2015saliency,koch1987shifts}, while others leverage machine learning techniques to estimate gaze position from indirect data: \citet{krafka2016eye} use webcam images, and \citet{son} and \citet{laconte2006predictive} use functional magnetic resonance imaging (fMRI). Similarly, \citet{o2018predicting} reconstructed fixations maps, which can predict eye movement patterns, directly from fMRI data.
Interestingly, our recent work indicates that it might be possible to infer gaze direction directly from EEG data \cite{10.1145/3448018.3458014}.
%Recently, in order to improve existing eye-tracking systems, studies have focused on leveraging advanced machine-learning techniques to compute gaze on indirect data, 
%such as webcam images \cite{krafka2016eye} or brain measures, e.g., derived from fMRI \cite{son,laconte2006predictive}. These works, reported significant advancements, such as the reconstruction of fixation maps, which can predict eye movement patterns from fMRI activity directly \cite{o2018predicting}. 
%Furthermore, there has been research carried out on the classic supervised machine learning techniques; for example, \cite{eth} proposed a new method for assessing repetitive patterns of eye movements and demonstrated the possibility of eye-based activity recognition. 
%Interestingly, our recent work indicates that it might be possible to infer gaze direction directly from EEG data \cite{10.1145/3448018.3458014}.
Although to our knowledge no previous work has employed deep learning to estimate eye position from EEG, there are studies demonstrating that combining EEG and ET can improve performance in various tasks, as compared with a single modality. In particular, this has been reported in the vigilance estimation~\cite{zheng2017multimodal}, information extraction and sentiment analysis \cite{hollenstein2018zuco}, or analysing users’ behaviour when performing a web search \cite{slanzi2017combining}. %in order to predict click intentions \cite{slanzi2017combining}.

\paragraph{Related Datasets.}%\
%Current openly available datasets that combine EEG and ET data encompass a wide range of possible scientific analyses and methods development. 
There exist some openly available datasets that combine EEG and ET data. However, in comparison to EEGEyeNet, they are acquired from a smaller sample of individuals \cite{hollenstein2018zuco, zuco2} or using a less advanced EEG-ET setup \cite{notaro2018simultaneous, nikolopoulos2017multimodal}.
A prominent example of a multimodal neurophysiological dataset, including EEG and ET data collected from a significant sample of participants (126 individuals), was proposed by \citet{langer2017resource}. However, it is devised as a resource for assessing information processing in the developing brain and contains data from young participants only. This way, EEGEyeNet dataset is the first large-scale and precisely annotated EEG-ET dataset containing recording from participant across the adult lifespan (18-80 years old). 
In Table \ref{tab:datasets:comparison} we compare the EEGEyeNet dataset with the cited datasets in terms of the number of participants, their age, and recording/session length.

\begin{table*}[h]
\centering
\scalebox{.95}{
\begin{tabular}{@{}l*{7}{S[table-format=-3.4]}@{}}
\toprule
{Paper/Dataset}  & {$Participants$} & {$Female / Male$} & {$Age$} & {$Recording$ $Length$} & {$Session$ $Length$} \\
\midrule  
{ZuCo 1.0}  & \makebox{12} & \makebox{5/7} & \makebox{25-51} & \makebox{x} & \makebox{48-72h}
\\
{ZuCo 2.0}  & \makebox{18} & \makebox{8/10} & \makebox{23-52} & \makebox{x} & \makebox{30h-54h}
\\
{Notaro et al} & \makebox{22} & \makebox{11/11}  & \makebox{18-40} & \makebox{7h}  & \makebox{x}
\\
{MAMEM} & \makebox{36} & \makebox{9/27}  & \makebox{25-71} & \makebox{x}  & \makebox{52h}
\\
{Langer et al}  & \makebox{126} & \makebox{56/70} & \makebox{6-44} &  \makebox{x} & \makebox{378h}
\\
{EEGEyeNet} & \makebox{356} & \makebox{190/166} & \makebox{18-80} & \makebox{47h} & \makebox{415h}
\\
%& \makebox{0 \rpm 0.0} & \makebox{0 \rpm 0.0} & \makebox{0 \rpm 0.0} & \makebox{0 \rpm 0} \\   
\bottomrule
\end{tabular}
}
\caption{Metadata comparison of the cited datasets.
Since \cite{hollenstein2018zuco, zuco2, nikolopoulos2017multimodal,langer2017resource} reported only the duration of the entire experimental session (including participant's preparation, practice trials that were aimed to acquaint the participant with the experimental procedures, breaks between the subsequent experiments, etc.), we decided to distinguish the recording's length from the whole session length.}  
\label{tab:datasets:comparison}
\end{table*}

\section{EEGEyeNet Dataset}\label{sec:dataset}

In this section, we provide a detailed description of the EEGEyeNet dataset.
Together with the raw data, we release two sets of preprocessed data: minimally and maximally preprocessed; as well as the preprocessing code. 
This way, we give users the freedom to manipulate raw data while easing the experimentation barrier by additionally providing ready-to-use clean data. 
%We emphasize that this dataset constitutes an independent contribution since it may be used for a variety of research purposes beyond evaluation on the benchmark presented in Section \ref{sec:benchmark}.

\subsection{Data Acquisition}\label{sec:dataset2}
\paragraph{Participants.}
Data were recorded from 356 healthy adults. The study included 190 female and 166 male participants, of ages between 18 and 80 years. All participants gave their written informed consent before participation in the experiment and received a monetary compensation (the local currency equivalent of 50 US Dollars). The data was collected according to the principles expressed in the Declaration of Helsinki \cite{world2013world}.

%old : {'m': 91, 'w': 111}, age mean (years) : 68.63, age std : 5.73
%young : {'m': 55, 'w': 110}, age mean (years) : 24.97, age std : 5.05

%\paragraph{Recording Setup.}%Electroencephlography and Eye-Tracking Aquisition}
%For the whole dataset, the identical acquisition setup has been used. The experimental setup reflects the current state of the art. 
\begin{wrapfigure}{r}{0.58\linewidth}
\centering
 \includegraphics[width=\linewidth]{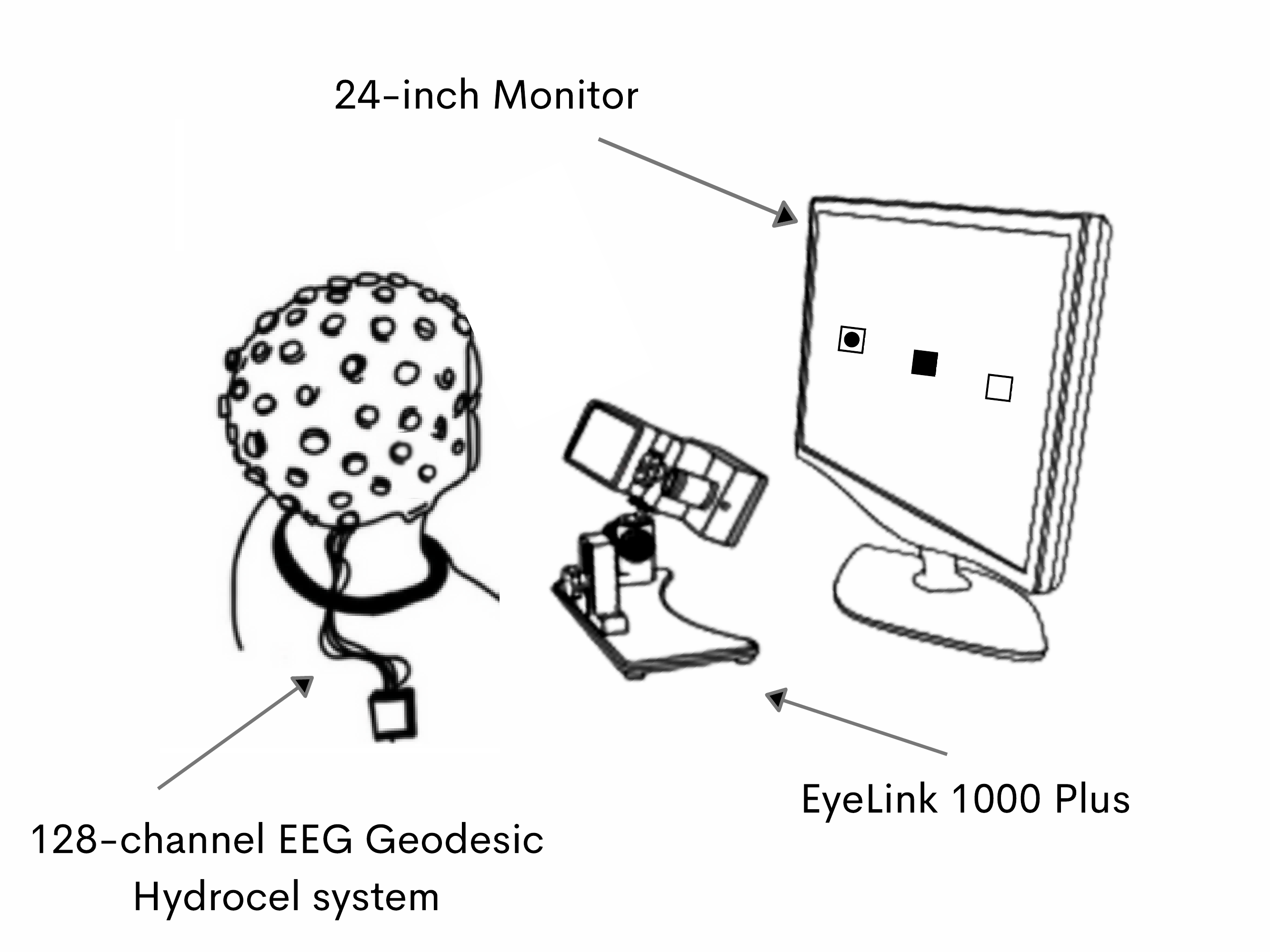}
  \caption{Recording Setup}
  \label{fig:recording-setup}
  \end{wrapfigure}
\paragraph{Recording setup.} High-density EEG data was recorded at a sampling rate of 500 Hz, with midline central recording reference, using a 128-channel EEG Geodesic Hydrocel system. The impedance of each electrode was checked prior to each recording session and kept below 40 kOhm.
Simultaneously, eye position was recorded with an infrared video-based ET EyeLink 1000 Plus from SR Research at a sampling rate of 500 Hz and an instrument spatial resolution of less than 0.01$\degree$ root mean square (RMS) of the distances between successive samples. The ET was calibrated with a 9-point grid before each recording. In a validation step, the ET calibration was repeated until the error between two measurements at any point was less than 0.5$\degree$, or the average error for all points was less than 1$\degree$.
Participants were seated at a distance of 68 cm from a 24-inch monitor with a resolution of $800 \times 600$ pixels.
A stable head position was ensured with a chin rest. The illustration of the recording setup can be seen in Figure \ref{fig:recording-setup}.

\subsection{Preprocessing}
EEG data is often contaminated by artifacts produced by environmental factors, e.g., temperature, air humidity, as well as other sources of electromagnetic noise, such as line noise~\cite{kappenman2010effects}.
These artifacts interact in a complex manner with participant-related artifacts, typically reflecting unwanted physiological signals such as eye movements, eye blinks, muscular noise, heart signals or sweating, which differ from participant to participant. The resulting artifacts in the EEG data are typically more prominent than the signal of interest (i.e. brain activity). Therefore, EEG data requires preprocessing in the form of artifact cleaning or artifact correction \cite{keil2014committee}. We preprocessed the entire EEGEyeNet dataset using the openly available toolbox from \citet{ped} in two ways: minimally and maximally. Minimal preprocessing includes the detection and interpolation of bad electrodes, and filtering the data with 40 Hz high-pass filter and 0.5 Hz low-pass filter (cf. Appendix \ref{app:prep}). The difference between these two types of preprocessing is that maximal preprocessing removes a much larger number of artifacts (muscles, heart, eyes, line noise, channel noise). To do this, independent component analysis (ICA) is applied in combination with \emph{IClabel}~\cite{pion2019iclabel}, a pre-trained classifier that estimates the probability of a component reflecting artifactual activity. If a component receives a probability estimation larger than $0.8$ for any class of artifact we remove it from the data. Minimally preprocessed data includes ocular artifacts, which is expected to make the estimation of gaze position easier. On the other hand, maximal preprocessing is a state-of-the-art technique for neuroscientific applications \cite{ped} that aims to keep only neurophysiological information in the data.

%EEG data is often contaminated by artifacts produced by technical and environmental factors (e.g., temperature, air humidity) as well as by the recorded participant. Additionally, these factors interact with other sources of noise, such as line noise or other electromagnetic noise \cite{kappenman2010effects}. Moreover, participant-related artifacts, typically reflecting unwanted physiological signals (such as eye movements, eye blinks, muscular noise, heart signals and sweating), may differ from participant to participant and interact in a complex manner with non-physiological artifacts. The resulting artifact signals in the EEG are typically more prominent than the signal of interest (i.e. brain activity). They therefore require preprocessing in the form of artifact cleaning or artifact correction \cite{keil2014committee}.

%This toolbox includes state-of-the-art EEG preprocessing methods and allows us to objectively preprocess the EEG data and quantify the quality of the preprocessed data~\cite{ped}. 
%EEG measurements are preprocessed in two ways: minimally and maximally. 

 %This way, EEGEyeNet is already prepared for both use cases, cognitive and behavioral research as well as EEG-based gaze estimation. %Note that for the same reasons, in our proposed benchmark, we use only minimally preprocessed data.

After preprocessing (both minimally and maximally), the EEG and eye-tracking data were synchronized using ``EYE EEG'' \cite{dimigen2011coregistration} to enable EEG analyses time-locked to the onsets of relevant events
%saccades, fixations, and other triggers, 
depending on the experimental paradigm. Synchronization quality was ensured by comparing the trigger latencies recorded in the EEG and eye-tracker data. All synchronization errors did not exceed 2 ms.

\subsection{Data Annotation}\label{sec:annot}
Existing literature studying eye movement generally distinguishes between three different events \cite{toivanen2015probabilistic}: saccades, fixations, and blinks. \textbf{Saccades} are rapid, ballistic eye movements that instantly change the gaze position. \textbf{Fixations} are defined as time periods without saccades, and \textbf{blinks} are considered a special case of fixation, where the pupil diameter is zero.
For each of the experimental paradigms described in Section \ref{sec:para}, we provide annotations in the form of start and end time of each event, as well as the start and end position of saccades and the average position of fixations. 
%in the form of a data structure that contains all task-related event markers together with the extracted saccades, fixations and blinks. 
%For each saccade we provide start and end time as well as start and end gaze position. For each fixation, we provide start and end time, and the average position of the gaze. Finally, we provide start and end time of each blink. 
See Appendix~\ref{app:ann} for further details.

\subsection{Experimental Paradigms}\label{sec:para}
%providing data for current benchmark task, i.e., predicting gaze position and eye movement direction from EEG data.

% \subsection{Figures}

%\begin{figure}
%  \centering
  %\fbox{\rule[-.5cm]{0cm}{4cm} \rule[-.5cm]{4cm}{0cm}}
%  \includegraphics[width=10cm]{ok.pdf}
%  \caption{Sample figure caption.}
%\end{figure}       

%\begin{figure}[!t]
%    \begin{minipage}{0.33\linewidth}
%      \includegraphics[width=1.1\textwidth]{AS.pdf}
%      \caption{}\label{fig:proanti}
%    \end{minipage}
%    \hfill
%    \begin{minipage}{0.32\linewidth}
%      \includegraphics[width=1.1\textwidth]{LG.pdf}
%      \caption{}\label{fig:largegrid}
%    \end{minipage}
%    \begin{minipage}{0.33\linewidth}
%      \includegraphics[width=1.1\textwidth]{VSST.pdf}
%      \caption{}\label{fig:vss}
%    \end{minipage}
%  \end{figure}

% \begin{wrapfigure}{r}{0.5\textwidth}
% \begin{center}
%     \includegraphics[width=0.48\textwidth]{AStasksep.pdf}
%   \end{center}
%   \caption{A caption}
% \end{wrapfigure}

% \noindent\begin{minipage}{0.6\textwidth}
% \includegraphics[width=\linewidth]{AStasksep.pdf}
% \end{minipage}%
% \hfill%

%\begin{minipage}{0.4\textwidth}
\begin{wrapfigure}{r}{0.5\linewidth}

\vspace*{-0.5cm}
\centering
\includegraphics[width=0.5\textwidth]{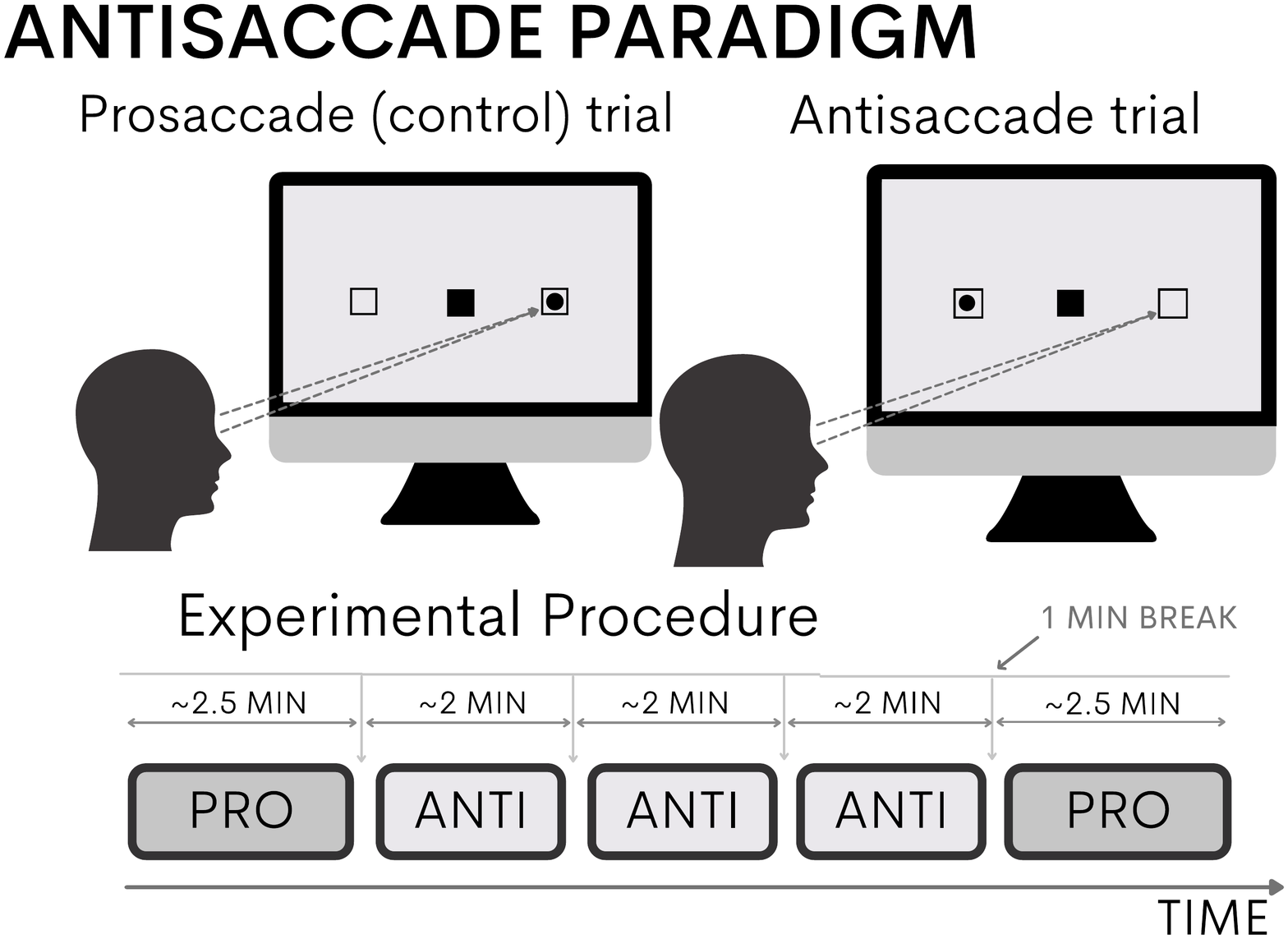}
\caption{ \textbf{Pro- and Antisaccade.} Schematic view of the experimental setup and gaze behavior during a prosaccade (left) and antisaccade (right) trial. }\label{fig:proanti}
\end{wrapfigure}
\paragraph{Pro- and Antisaccade.}
The pro- and antisaccade paradigm was designed according to the internationally standardized protocol for antisaccade testing developed by \citet{antoniades2013internationally} and is described in detail in \cite{martynka}. Each trial starts with a central fixation square. The participants are asked to focus on the center of the screen for a randomized time-period between 1 and 3.5 seconds. Subsequently, the cue (i.e. a dot) appears horizontally on the left or the right hand-side of the central fixation square. In the prosaccade trials, the participants are asked to focus their gaze on the cue as fast as possible, while in the antisaccade trials the participants are instructed to perform a saccade towards the opposite side of the cue. In both cases, the cue is shown for one second. As soon as the cue disappears, the participants shift their focus back to the center of the screen; this is illustrated in Figure~\ref{fig:proanti}. Data recorded following this paradigm may be used for different research purposes, such as estimating gaze direction or examining responses to inhibition. %brain activity during inhibitory control. 

%\end{minipage}

% \begin{wrapfigure}{r}{0.5\textwidth}
% \begin{center}
%     \includegraphics[width=0.48\textwidth]{vistasksep.pdf}
%   \end{center}
%   \caption{A caption}
% \end{wrapfigure}

% \noindent\begin{minipage}{0.6\textwidth}
% \includegraphics[width=\linewidth]{vistasksep.pdf}
% \end{minipage}%
% \hfill%
% \begin{minipage}{0.4\textwidth}

% \end{minipage}

% \begin{wrapfigure}{r}{0.5\textwidth}
% \begin{center}
%     \includegraphics[width=0.48\textwidth]{largsep.pdf}
%   \end{center}
%   \caption{A caption}
% \end{wrapfigure}

% \noindent\begin{minipage}{0.6\textwidth}
% \includegraphics[width=\linewidth]{largsep.pdf}
% \end{minipage}%
% \hfill%
% \begin{minipage}{0.4\textwidth}
\begin{wrapfigure}{r}{0.5\linewidth}
\centering
\includegraphics[width=0.5\textwidth]{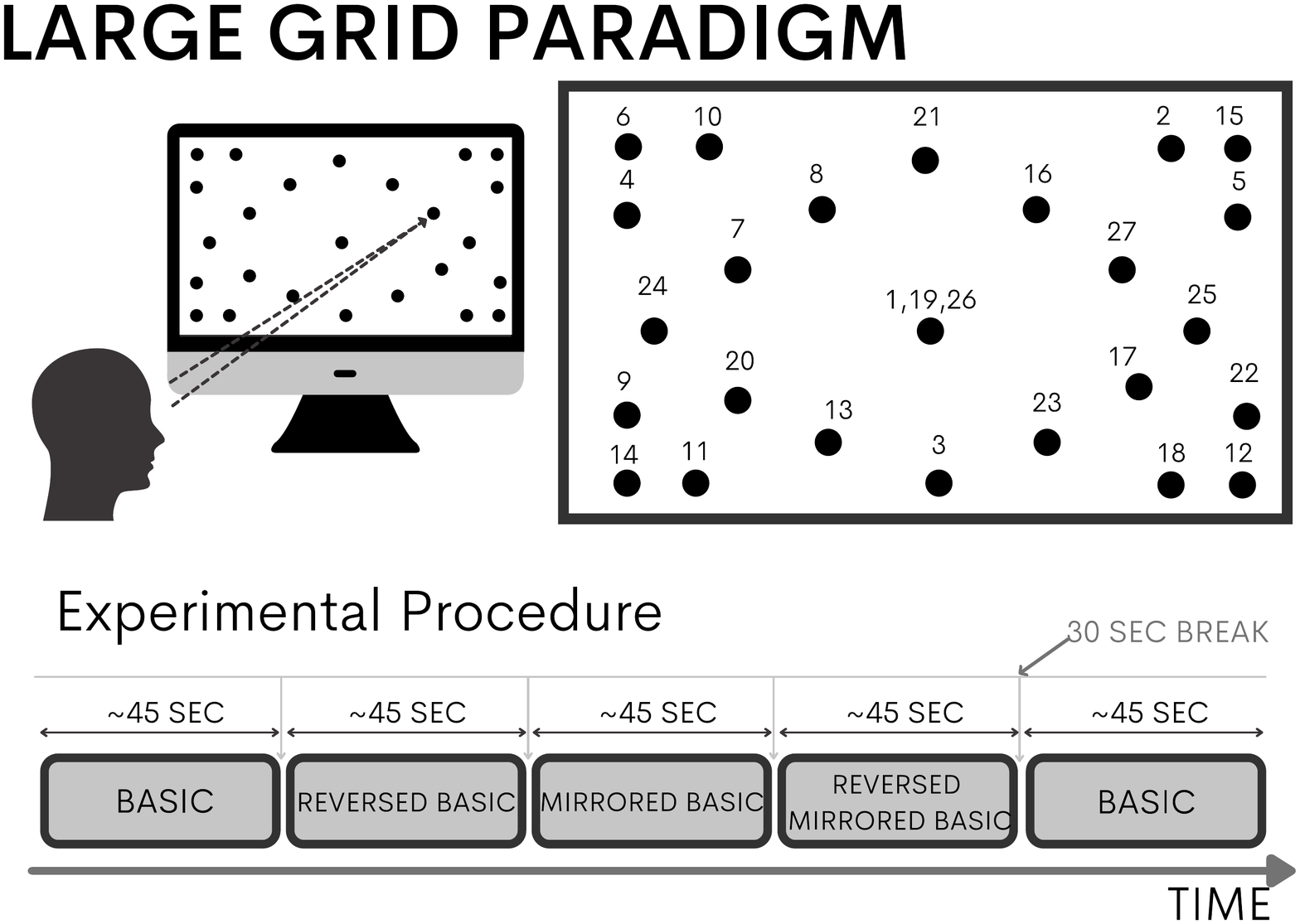}
\caption{\textbf{Large Grid.} Schematic view of the experimental setup and stimuli location on the screen. %during the Large Grid Paradigm.
}\label{fig:largegrid}
\end{wrapfigure}
\paragraph{Large Grid.}
Participants are asked to fixate on a series of dots that are sequentially presented, each at one of 25 different screen positions. Unlike the others, the dot at the center of the screen appears three times, resulting in 27 trials (displayed dots) per block, each dot is displayed for $1.5$ to $1.8$ seconds. The positions of the dots were selected to ensure coverage of all corners of the screen as well as the center (see Figure~\ref{fig:largegrid}). The shape of the grid and its use for eye gaze estimation follows the work from \citet{son}. Given that \citet{son} used the Large Grid paradigm for functional Magnetic Resonance Imaging (fMRI), we have adapted the length of the stimulus and the number of repetitions to our setup. To record a larger number of trials and reduce the predictability of the subsequent positions in the primary sequence of the stimulus, we use different pseudo-randomized orderings of the dots presentation, distributed in five experimental blocks, as shown in Figure~\ref{fig:largegrid}.
%we use different pseudo-randomized orderings of the dots presentation, distributed in five experimental blocks, as shown in Figure~\ref{fig:largegrid}.
%Each trial on the Large Grid Paradigm contains five blocks. The first and fifth blocks' order (``basic'') is presented in Figure 1. In the second block, the ``basic'' order is reversed; then, in the third and fourth block, the ``basic'' order is mirrored and reversed mirrored, respectively.
The entire procedure is repeated 6 times during the measurement, resulting in 810 stimuli for each participant.
%\end{minipage}

%. The appearance of the dots in the left- and the right-hand square
%was randomized. The subtasks were presented in blocks in the following order: 1. Prosaccade, 2. Antisaccade, 3. Antisaccade, 4. Antisaccade, 5. Prosaccade. Each prosaccade
%block consisted of 60 trials (30 trials per visual hemifield) and each anti-saccade block of 40
%trials (20 trials per visual hemifield). There was a one-minute break between each block.
\begin{wrapfigure}{r}{0.5\linewidth}
\centering
\includegraphics[width=0.5\textwidth]{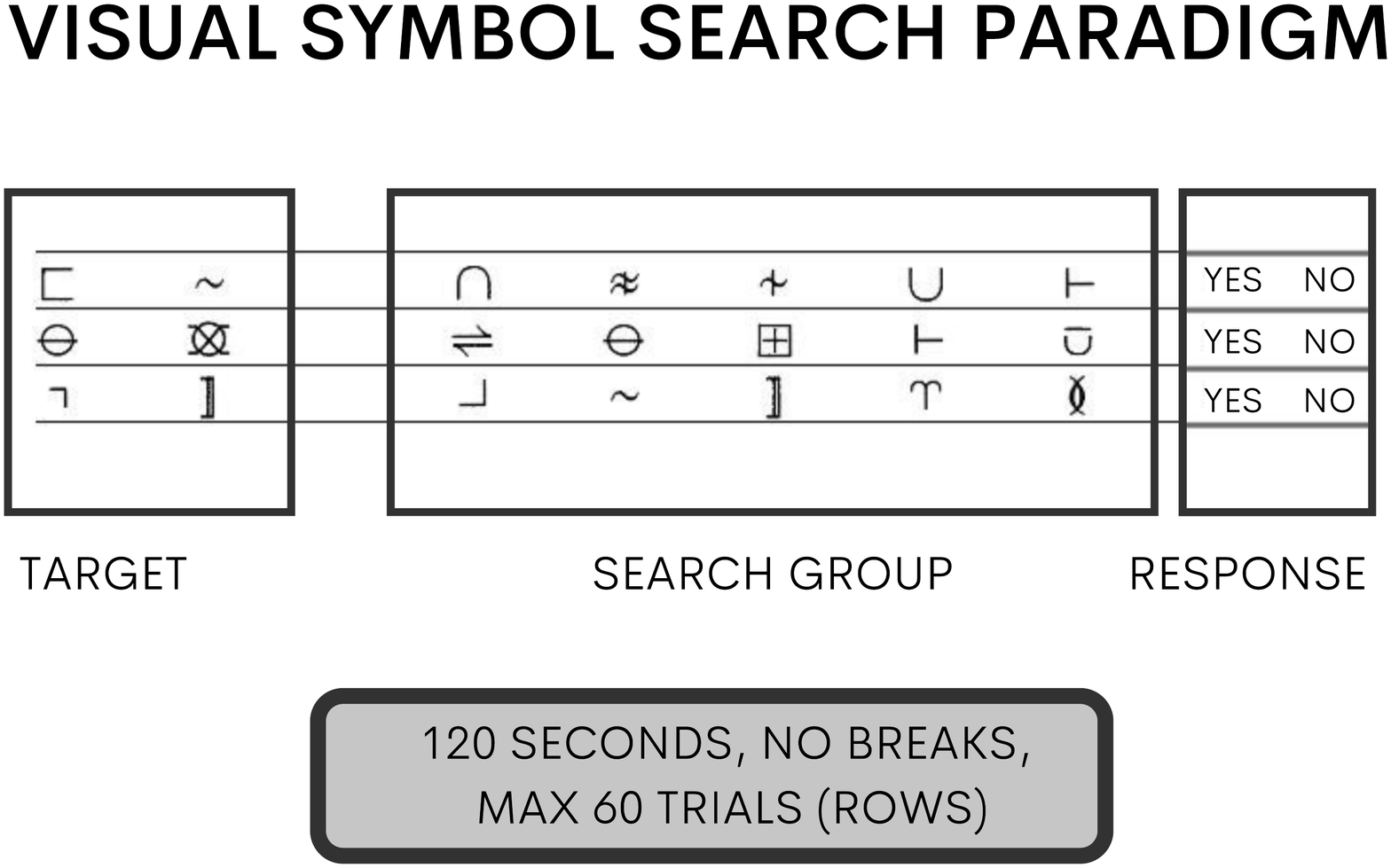}
\caption{\textbf{Visual Symbol Search.}
Three representative trials of the VSS task. For each row, the first two symbols are target symbols, the next five correspond to the search group, and the response check boxes are labeled as response.}\label{fig:vss}
\end{wrapfigure}
\paragraph{Visual Symbol Search.}
%Here we should only mention the nature of the paradigm, and how it could be useful for different tasks (but not explain the benchmarking tasks here). So for example, for antisaccade we can argue that because of the nature of the paradigm, the eye movements are predictable/exact/ between 2 points (left and right). And this makes it a perfect candidate to create a benchmark for predicting left and right. For visual search, we can say typically people look in the same line while searching the symbol. That means the eye movements found here are also mostly between left and right, but with different lengths/amplitudes.....
Visual Symbol Search (VSS) is a computerized version of a clinical assessment to measure processing speed (Symbol Search Subtest of the Wechsler Intelligence Scale for Children IV (WISC-IV)~\cite{wechsler1949wechsler}) and the Wechsler Adult Intelligence Scale (WAIS-III) ~\cite{wechsler1955wechsler}). Participants are shown 15 rows at a time, where each row consists of two \emph{target} symbols, five \emph{search} symbols and two additional symbols that contain respectively the words ``YES'' and ``NO''. For each row, participants need to indicate by clicking with the mouse button on the ``YES'' or ``NO'' symbol, whether or not one of the two target symbols appears among the five search symbols. Each recording of the VSS paradigm takes 120 seconds with a maximum of 60 trials, where one trial corresponds to one row; in $50\%$ of the trials one of the target symbols does appear in the search symbols and in the remaining $50\%$ none does.
Once participants finish a set of 15 rows, they press a ``next page'' button which displays a new set of 15 rows. Participants are instructed to solve as many rows, or trials, as possible within the given 120 seconds.
Before beginning the actual recording, participants perform a training of four trials, for which they receive feedback to ensure they understand the task. No feedback is provided throughout the actual recording. Data collected according to this paradigm may be used for investigating behavioral and neurophysiological correlates of processing speed. % be used to THIS AND THAT.
%Once a participant finishes a set of 15 rows, they press a ``next page'' button which displays a new set of 15 rows. There was maximum of 60 trials in total. Participants were instructed to solve as many rows, or trials, as possible within 120 seconds.

The EEGEyeNet dataset contains data recorded following all three different experimental paradigms mentioned above. These paradigms cover typical cognitive tasks and provide a wealth of eye movement patterns.

%%%%%%%%%%%%%%%%%%%%%%%%%%%%%%%%%%%%%

%\paragraph{Eye tracking Preprocessing}
%The algorithm provided by EyeLink 1000 Plus was used to identify saccades, fixations and blinks. The acceleration threshold was set as 8000° per second, the velocity threshold as 30° per second and the deflection threshold as 0.1°. The resulting data set included gaze locations in a XY-coordinate system with pixels as units 

%\subsection{Event Summary}

Finally, in Table~\ref{tab:summary} we report for each paradigm the number of appearances of the three events that we extract: fixations, saccades and blinks; we report these after both minimal (min) and maximal (max) preprocessing. In Appendix \ref{app:sc} we give further details about the characteristics of each of these events. These numbers make apparent the large size of the EEGEyeNet dataset, with a total of over 47 hours of recorded events.

%this is in the preparation

%\begin{figure}[!t]
 %   
  %    \includegraphics[width=0.99\textwidth]{Screenshot 2021-06-04 at 10.04.58.png}
   %   \caption{Dummy}\label{fig:proanti}
  %\end{figure}

\begin{table*}[h]
\centering
\begin{tabular}{@{}l*{8}{S[table-format=-3.4]}@{}}
\toprule
{Paradigm} & {Preproc.} & {\# Fixations} & {\# Saccades} & {\# Blinks}  & {Total time} \\
\midrule
\multirow{2}{*}{\emph{Pro- Antisac.}} & {min} & {357115} & {358384} & {56179} & {38 h} \\ %{136776746} \\
& {max} & {358587} & {359856} & {57991}  & {38 h 6 mins} \\ %{137199492} \\
\midrule
\multirow{2}{*}{\emph{Large Grid}} & {min} & {68075} & {68245} & {11108}  & {7 h 52min} \\ %{28373692} \\
& {max} & {69013} & {69185} & {11237} & {7 h 58 min} \\ %{28696957} \\
\midrule
\multirow{2}{*}{\emph{VSS}} & {min} & {43384} & {43443}  & {971}   & {1 h 29 min} \\ %{5346982} \\
& {max} & {43279} & {43339}  & {945}  & {1 h 28 min}\\
\midrule
\multirow{2}{*}{Total} & {min} & {468574} & {470072} & {68258}  &  {47 h 21 min} \\ %{170497420}\\
& {max} & {470879} & {472380} & {70173} &  {47 h 33 min} \\ %{171193822}\\
\bottomrule
%& \makebox{0 \rpm 0.0} & \makebox{0 \rpm 0.0} & \makebox{0 \rpm 0.0} & \makebox{0 \rpm 0} \\   
\bottomrule
\end{tabular}
\caption{ \textbf{Overview of EEGEyeNet Dataset.} Number of eye-tracking events (fixations, saccades, blinks) for the EEGEyeNet experimental paradigms.
The difference in the number of eye events between the two preprocessing versions is due to fact that in the minimal preprocessing more events are identified as artifacts and removed from the sample.
}  
\label{tab:summary}
\end{table*}  

% \begin{table*}[h]
% \centering
% \resizebox{\textwidth}{!}{\begin{tabular}{@{}l*{8}{S[table-format=-3.4]}@{}}
% \toprule
% {Paradigm} & {Preproc.} & {\# Saccades} & {\# Blinks} & {\# Fixations} & {Total time [ms]} \\
% \midrule
% \multirow{2}{*}{\emph{Pro- Antisac.}} & {min} & {358384} & {56179} & {357115} & {136776746} \\
% & {max} & {359856} & {57991}  & {358587} & {137199492} \\
% \midrule
% \multirow{2}{*}{\emph{Large Grid}} & {min} & {68245} & {11108}  & {68075} & {28373692} \\
% & {max} & {69185} & {11237} & {69013} & {28696957} \\
% \midrule
% \multirow{2}{*}{\emph{VSS}} & {min} & {43443}  & {971}  & {43384}  & {5346982} \\
% & {max} & {43339}  & {945}  & {43279} & {5297373}\\
% \midrule
% \multirow{2}{*}{Total} & {min} & {470072} & {68258} & {468574} &  {170497420}\\
% & {max} & {472380} & {70173} & {470879} &  {171193822}\\
% \bottomrule
% %& \makebox{0 \rpm 0.0} & \makebox{0 \rpm 0.0} & \makebox{0 \rpm 0.0} & \makebox{0 \rpm 0} \\   
% \bottomrule
% \end{tabular}}
% \caption{ Datasets Statistics
% }  
% \label{tab:summary}
% \end{table*}  

%total time converted
%anti_max : 76h13mn18s984ms
%anti_min : 75h59mn13s492ms
%dots_max : 15h56mn33s914ms
%dots_min : 15h45mn47s384ms
%prospeed_max : 2h56mn34s746ms
%prospeed_min : 2h58mn13s964ms

%\subsection{Data Preparation}

%%%%%%%%%%%%%%%%%%%%%%%%%%%%%%%%%%%%%%%%%%%%%%%%%%%%%%%%%%%%%%%%%%%%%%%%%%%%%%%%%%%%%%%%% 

\section{Benchmark}
\label{sec:benchmark}

 \begin{figure}[b]
  \centering
 \includegraphics[width=0.65\linewidth]{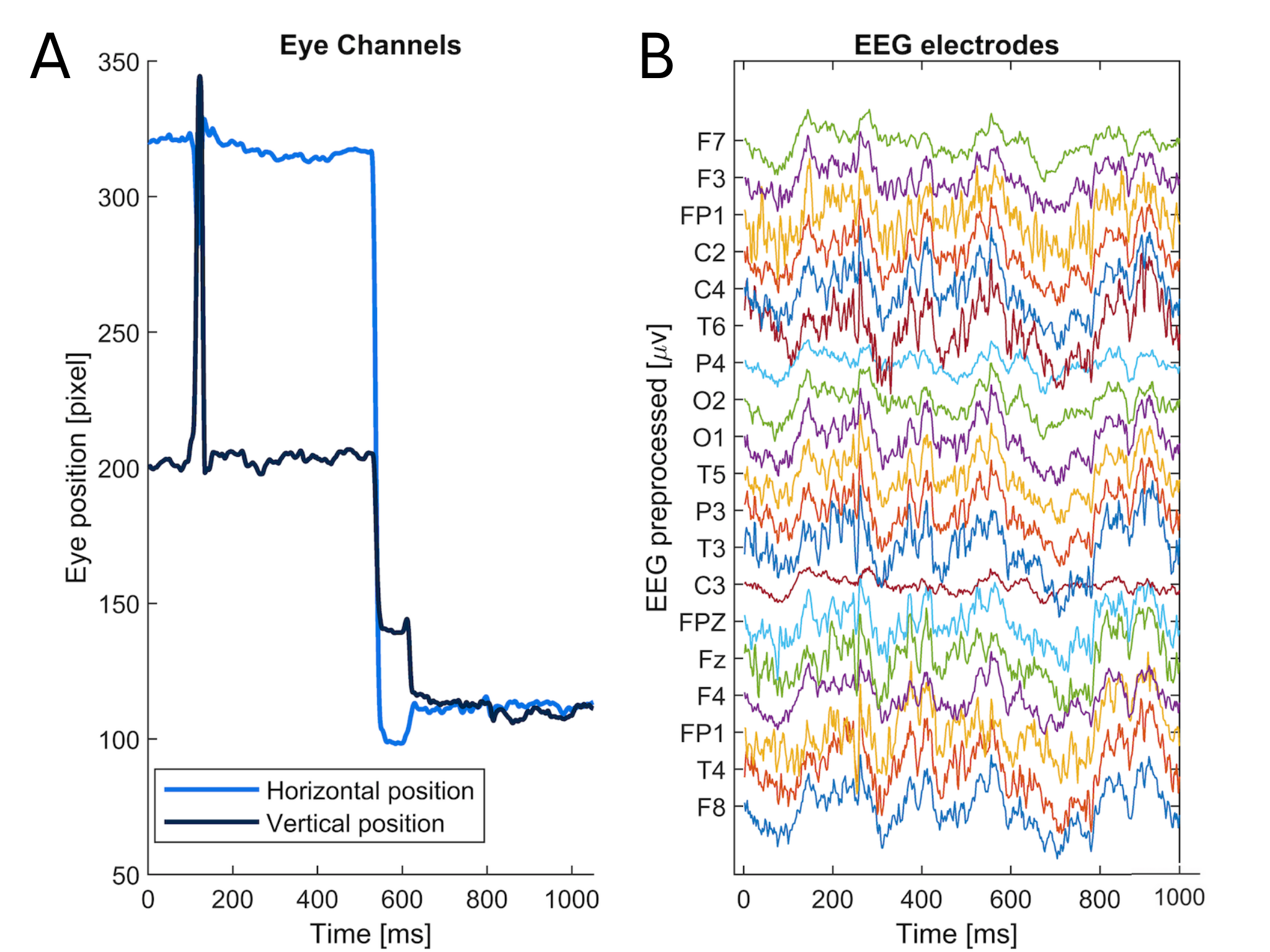}
  \caption{ Each sample of EEG data has a shape of $500 \times 128$, i.e., $500$ time points for each of the $128$ EEG channels. On the left side (A) we can see gaze data (along \emph{XY}-axes) of the one-second sample. On the right side (B) we can see a subset of the preprocessed EEG data (electrodes matching the 10–20 systems were chosen). }
  \label{fig:sample}
  \end{figure}

Based on the EEGEyeNet dataset we propose a benchmark to assess EEG-based eye tracking methods. This benchmark %considers a subset of data from the complete EEGEyeNet dataset and 
consists of three different tasks with increasing difficulty. For each of the tasks, we provide data preparation modules that cut the data into samples of one second with a temporal resolution of $2 ms$ from all 128 EEG channels; This way all samples have a shape of $500 \times 128$, i.e., $500$ time points for each of the $128$ EEG channels. This is also illustrated in Figure \ref{fig:sample}. We use exclusively minimally preprocessed EEG data as it produces better performance as compared to maximal preprocessed data. In Appendix~\ref{app:FurthExp}, we report our results for the maximally preprocessed data, as well as the results of the experiments ran on the Zuco 2.0 dataset \cite{zuco2}. 
For comparability, we also provide a stable split of the data across participants, with 70\% of the participants for training, 15\% for validation and 15\% for test. Note that each participant's data is contained only in one of the three sets, i.e., the same participant does not appear during training and validation or test. In Table~\ref{tab:benStats} we include a summary of how the data is split in each task of the benchmark.

\begin{table*}[h]
\centering
\resizebox{\textwidth}{!}{\begin{tabular}{@{}lc*{9}{S[table-format=-3.4]}@{}}
\toprule
& \multicolumn{4}{c}{\# Participants} & \multicolumn{4}{c}{\# Samples} & \\
\cmidrule(lr){2-5} \cmidrule(lr){6-9}
{Dataset} & {Total} & {Train} & {Validation} & {Test} & {Total} & {Train} & {Validation} & {Test} \\
\midrule
%{lrantimax} & 330 & 230 & 50 & 50 & 30825 & 21061 & 4946 & 4818 \\
{Left-Right} & {329} & {229} & {50} & {50} & {30842} & {21042} & {4980} & {4820} \\
%{dirdotsmax} & 19 & 13 & 3 & 3 & 10896 & 7356 & 1877 & 1663 \\
{Angle/Amplitude} & {27} & {19} & {4} & {4} & {17830} & {12275} & {2836} & {2719} \\
%{dirprospeedmax} & 86 & 60 & 13 & 13 & 31191 & 21529 & 4676 & 4986\\
%{dirprospeedmin} & 87 & 59 & 14 & 14 & 31563 & 21191 & 5018 & 5354\\
%{posdotsmax} & 19 & 13 & 3 & 3 & 13561 & 9243 & 2137 & 2181 \\
{Abs. Position} & {27} & {19} & {4} & {4} & {21464} & {14706} & {3277} & {3481} \\
%& \makebox{0 \rpm 0.0} & \makebox{0 \rpm 0.0} & \makebox{0 \rpm 0.0} & \makebox{0 \rpm 0} \\   
\bottomrule
\end{tabular}}
\caption{\textbf{Benchmark Data.} Detail of the data split for each task of the proposed benchmark.
}  
\label{tab:benStats}
\end{table*}

\subsection{Task 1: Left-right}

The first and simplest task of our benchmark consists of determining the direction of the subject's gaze along the horizontal axis. This task is performed exclusively on the first experimental paradigm (\emph{Pro- and Antisaccade}). 
We make sure that in all the samples provided for this task, after the cue, the participant performs a saccade towards the correct direction and then fixates on the cue; otherwise, the sample is not included in the data. %Furthermore, in this task, we use data only from the prosaccade block of the antisaccade experimental paradigm.
%ntisaccades give a measure of inhibition control since it requires inhibition of the reflexive response towards the direction of the cue and a voluntary movement in the opposite direction \cite{antoniades2013internationally}. 
%The antisaccade experimental paradigm is a measure of inhibition control since it requires inhibition of the reflexive response in the direction of the cue presentation and a voluntary movement in the opposite direction \cite{antoniades2013internationally}. 
Given that the antisaccade task has been viewed as an indicator of behavioral inhibition abilities \cite{klimesch2007eeg} and that here we are interested in saccade direction, for the sake of simplicity, we use only data from the prosaccade trials of the antisaccade experimental paradigm. %To encourage experimenation the data from the antisaccade condition is released as well.
From the $30,842$ samples used in this task, $14,827$ correspond to the ``left'' label and $16,015$ to ``right''.

%The data for this task is provided as pre-processed samples of one second of duration with a temporal resolution of $2 ms$ from all the 128 EEG channels. This way the samples have a shape of $500 \times 128$, i.e., $500$ time points for each of the $128$ EEG channels. 
%In this task, each sample starts at the instant when a cue is shown on the screen and their length is set to one second because in the the antisaccade paradigm the time between cues is at least one second.
Each sample starts at the instant when a cue is shown on the screen and contains one saccade: the goal is to determine whether it is a left or a right saccade. This is a binary classification task, and therefore, performance is measured in terms of accuracy with a random baseline of $52.3\%$, given by the majority class. Although seemingly simple, determining the horizontal direction of the human gaze is of paramount importance in multiple real-world applications, e.g., gaze-based writing systems for people with disabilities \cite{bates2007introducing}.

\subsection{Task 2: Angle/Amplitude}

The second task of our benchmark consists of determining the angle and amplitude of a saccade and is performed on data from the \emph{Large Grid} paradigm. 
%This task is performed on data from the second experimental paradigm (\emph{Large Grid}) and it includes recording from all the $19$ participants with a total of $17,830$ labelled samples, which are then split into train, validation and test. %from a given initial position. 
%Similar to the previous task, the sample length is fixed to one second and so, the sample shape is $500 \times 128$. 
The one-second samples in this task contain a saccade onset in the middle of the sample. This way, each sample contains a complete saccade as well as a part of the preceding and succeeding fixations. 
%Each sample consists of five events: fixation, end of the previous cue, next cue, saccade and fixation. The saccade is located in the middle of the sample and since the the fixations at the beginning and at the end are typically longer than $500ms$, only a part of these fixations is included. % in order to maintain a sample length constant $1$ second. 
During the data preparation for this task, we remove samples with fixations that are shorter than $150ms$ and with saccades with an amplitude of less than $1\degree$ \cite{martinez2013impact}. % degrees. 

Given a sample, the task of the model is to regress the two target values, i.e., angle and amplitude of the relative change of the gaze position during the saccade. The evaluation metric for this task is the square root of the mean squared error (RMSE) for the angle (in radians along the shortest path of the unit circle) and amplitude (in millimeters) separately. The naive baseline is given by the mean angle and amplitude in the training set and amounts to $1.90$ RMSE radians for the angle and $74.7$ RMSE mm for the amplitude.
This task is significantly harder than the Left-Right task and aims at serving as an intermediate step for the development of a purely EEG-based ET. Despite its difficulty, there is evidence that EEG recordings contain angle information \cite{hladek2018real}. 

%Similar to the previous task the evaluated model receives as input a sample of one second of EEG recording (128 channels) from which it should regress the two target values, i.e., angle and amplitude of the relative change of the gaze position. Each sample consists of a 5-tuple of events: fixation, end of the previous cue, next cue, saccade and fixation. The saccade is typically located in the middle of the sample. The first and the last fixation are typically longer than $500ms$, thus we only keep the EEG measurements of a part of the fixations to maintain a constant sample length of $1$ second. That is, the data preparation software tool cuts 500 datapoints, starting with the event where the participant is fixating towards a cue. Then, it checks whether in the mean time another cue is shown in the screen. Finally, it also checks that the participant performs a saccade and fixation afterwards. The data preparation module also ignores fixations that are shorter than $150ms$ and saccades that have an amplitude of less than 2. The evaluation metric for this task is Mean Square Error (MSE) of both, angle and amplitude.

%This task is significantly harder than the previous one and aims at serving as an intermediate step for the development of a purely EEG-based ET. Despite its difficulty, there is evidence that EEG recordings contain angle information \cite{hladek2018real}. 

\subsection{Task 3: Absolute Position}\label{sec:absPos}

Finally, we propose the task of determining the absolute position of the subject's gaze in the screen, described in terms of \emph{XY}-coordinates. Again, this task is performed on the \emph{Large Grid} paradigm. We provide samples of one second during which the participant is performing only one fixation.
The data preparation module ensures that in this time window there is no other event happening, i.e., the participant performs only a fixation. % Before splitting, the dataset for this task is composed of labelled $21,464$ samples. 
However, we note that in order to estimate the current gaze position we expect past information, e.g., the previous gaze position, to be helpful. For this reason, we also encourage experimentation on different ways of processing and cutting the full EEG recordings provided in the dataset.

Performance is measured as the euclidean distance in millimeters between the actual and the estimated gaze position in the \emph{XY}-coordinate plane. Random performance is again calculated as the mean position across the training set and corresponds to a distance of $123.3$ mm. This is the hardest task in the proposed benchmark and aims at simulating a purely EEG-based eye-tracker. We expect this task to help in the development of general methods to further improve gaze estimation systems as well as in setting an upper bound on the reach of EEG-based eye-tracking.

%Here, the goal is to predict the position where the subject is fixating, and performance is measured as MSE in the XY coordinate plane. 
%In this task, we expect past information, i.e., previous gaze position, to help the model in determining the current position and so, we note that using only samples of one second may limit the ability of the models to estimate gaze position. For this reason, we also encourage experimentation on different ways of processing the full EEG recordings, which are also provided.
%here we provide the EEG recordings as a continuous stream of data from which the model should infer the location of the gaze at each point. This way, we expect past information, i.e., previous gaze position, to help the model in determining the current position.
%The performance in this task is measured as 

\section{Baselines} % TODO: think of better name

%lrantimax labels distribution - 0 : 14813 , 1 : 16012
%lrantimin labels distribution - 0 : 14827 , 1 : 16015

We run extensive experiments on the proposed benchmark in order to provide baselines of different complexity. In our repository we provide an intuitive interface to reproduce our results and to use the methods presented here as a starting point for further research. We consider both models based on classical machine learning as well as large neural networks.

\subsection{Models}

\paragraph{Machine Learning.} These models operate on features extracted from the data rather than on the raw data. Therefore, in a feature extraction step, we apply a band-pass filter in the alpha band [$8-13$ Hz] on the continuous EEG signals across the entire trial. The choice of the alpha band is motivated by growing evidence suggesting that alpha activity is integral to spatial attention, and therefore plays a central role in covert orienting in a range of paradigms \cite{foster2017alpha}. After band-passing the signal, the Hilbert transform was applied, resulting in a complex time series from which we extract phase and amplitude.
%The Hilbert phase and amplitude are extracted from the data, always locked to the saccade onset.
%to enable the power of this fre for time segments defined through the saccade onset from the eye-tracking recording. 
Using the resulting features, we experiment with different models. 
\emph{Left-Right} is a classification task while \emph{Angle/Amplitude} and \emph{Absolute Position} are regression tasks and thus, some of the considered models can be applied only to either of those two types of tasks. In particular, the classification-only models that we study are Gaussian Naive Bayes (NB) Linear Support Vector Classification (SVC) and Radial Basis Function (RBF) kernel SVC, whereas the regression-only models are Linear, Ridge and Lasso Regression as well as Elastic Net and RBF Support Vector Machine for regression (SVR). Furthermore, we use K-Nearest Neighbours (KNN) and four tree-based models, Random Forest, Gradient Boost, AdaBoost and XGBoost, which can be used for both, classification and regression. In all cases we use the implementation from the Sklearn library \cite{scikit-learn}, for detailed model hyperparameters see Appendix \ref{app:DL}.

\paragraph{Deep Learning.}
We evaluate five state-of-the-art architectures for time series on the proposed benchmark: a standard one-dimensional convolutional neural network (CNN), a CNN with pyramidal shape, the EEGNet model by \citet{eegnet}, an InceptionTime model~\cite{inceptiontime}, and an Xception model~\cite{xception}. All of these models use convolutions as the primary operation (see Appendix \ref{app:DL} for architectural details). We tune the learning rate and other hyperparameters on the validation set of the \emph{Left-Right} task and use those values in all the reported results (cf. Appendix \ref{app:DL}). We use binary cross entropy loss to train the models for \emph{Left-Right}, and mean square error (MSE) loss for the other two tasks; in all cases we use the Adam optimizer \cite{kingma2014adam} and early stopping on the validation sets.

% I put a / for the STD of models with closed form solution because it will always be zero
% I don't know the reason of the STD equals zero for the other models (implementation error ?)

\subsection{Results}

In Table~\ref{tab:results} we provide the results of our evaluation for each of the models considered and for each of the three tasks. 
%The deep learning models are evaluated as ensembles of XXXX models with the same architecture and different training seed, their prediction is decided by majority voting. 
We tune the learning rate of the different deep learning models on the validation set of the \emph{Left-Right} task and use the same rate in the other two tasks, i.e., $1e^{-4}$. We run 5 times each experiment and report mean performance and standard deviation. 

\begin{table*}[t]
\centering
\begin{tabular}{@{}lc*{5}{S[table-format=-3.4]}@{}}
\toprule
\toprule
& {Left-Right} & \multicolumn{2}{c}{Angle/Amplitude} & {Abs. Position}\\
\cmidrule(lr){2-2} \cmidrule(lr){3-4} \cmidrule(lr){5-5}
{Model} & {Accuracy} & {Angle RMSE} & {Amp. RMSE} & {RMSE} \\
\midrule  
{KNN} & \makebox{90.7 \rpm 0} & \makebox{1.26 \rpm 0} & \makebox{59.3 \rpm 0} & \makebox{119.7 \rpm 0}\\
{GaussianNB} & \makebox{87.7 \rpm 0} & \makebox{-} & \makebox{-} & \makebox{-}\\
{LinearSVC} & \makebox{92.0 \rpm 0} & \makebox{-} & \makebox{-} & \makebox{-}\\
{RBF SVC/SVR} & \makebox{89.4 \rpm 0} & \makebox{1.88 \rpm 0} & \makebox{75.9 \rpm 0} & \makebox{123 \rpm 0}\\
{Linear Regression} & \makebox{-} & \makebox{1.39 \rpm 0} & \makebox{64.6 \rpm 0} & \makebox{118.3 \rpm 0}\\
{Ridge Regression} & \makebox{-} & \makebox{1.39 \rpm 0} & \makebox{64.2 \rpm 0} & \makebox{118.2 \rpm 0}\\
{Lasso Regression} & \makebox{-} & \makebox{1.38 \rpm 0} & \makebox{63.9 \rpm 0} & \makebox{118 \rpm 0}\\
{Elastic Net} & \makebox{-} & \makebox{1.38 \rpm 0} & \makebox{63.6 \rpm 0} & \makebox{118.1 \rpm 0}\\
\midrule
%{Decision Tree} & \makebox{96.2 \rpm 0} & \makebox\\
{Random Forest} & \makebox{96.5 \rpm 0} & \makebox{1.09 \rpm 0.01} & \makebox{59.8 \rpm 0.1} & \makebox{116.7 \rpm 0.1}\\
{Gradient Boost} & \makebox{97.3 \rpm 0.1} & \makebox{1.11 \rpm 0.01} & \makebox{60 \rpm 0.1} & \makebox{117 \rpm 0.1}\\
{AdaBoost} & \makebox{96.3 \rpm 0} & \makebox{1.43 \rpm 0.01} & \makebox{65 \rpm 0.1} & \makebox{119.4 \rpm 0.1}\\
{XGBoost} & \makebox{97.9 \rpm 0} & \makebox{1.11 \rpm 0} & \makebox{61.3 \rpm 0} & \makebox{118 \rpm 0}\\
\midrule
{CNN} & \makebox{98.3 \rpm 0.5} & \makebox{\textbf{0.33} \rpm 0.05} & \makebox{32 \rpm 3.6} & \makebox{\textbf{70.2} \rpm 1.1}\\
{PyramidalCNN} & \makebox{98.5 \rpm 0.2}  & \makebox{0.34 \rpm 0.04} & \makebox{\textbf{30.7} \rpm 1} & \makebox{73.6 \rpm 1.9}\\
{EEGNet} & \makebox{98.6 \rpm 0.1}  & \makebox{0.70 \rpm 0.08} & \makebox{46 \rpm 5.2} &  \makebox{81.7 \rpm 1.0}\\
{InceptionTime} & \makebox{97.9 \rpm 1.1} & \makebox{0.44 \rpm 0.16} & \makebox{43.6 \rpm 21.85} & \makebox{70.8 \rpm 0.8}\\
{Xception} & \makebox{\textbf{98.8} \rpm 0.1} & \makebox{0.47 \rpm 0.28} & \makebox{32.2 \rpm 1.9} & \makebox{78.7 \rpm 1.6}\\
\midrule
{Naive Baseline} & \makebox{52.3} & \makebox{1.90} & \makebox{74.7} & \makebox{123.3} \\
\bottomrule
\end{tabular}
\caption{\textbf{Results.} Mean and standard deviation of 5 runs of the considered models on the three benchmark tasks. \emph{Angle} is measured in radians, \emph{Amplitude} and \emph{Abs. Position} in mm.
}  
\label{tab:results}
\end{table*}

\paragraph{Left-Right.} 
We see that classical machine learning models achieve high performance in this task, much above the naive baseline of $52.26\%$. In particular, tree-based models reach a performance of over $96\%$, which confirms Left-Right as the easiest task in the proposed benchmark. Notably, although classical models obtain high scores, all the deep learning models reach a performance consistently higher, with an accuracy of over $98\%$ in all cases (except for InceptionTime, $97.9\%)$. This shows that despite the high performance, differences in performance can still be noticeable. 
% reason for the high performance in this task, is the large amount of data that we make available in our dataset, which allows the models to fully exploit their expressive capacity, unlike in related datasets, where the best performing models reached an accuracy of only $X\%$. 
% Sanity check?

\paragraph{Angle/Amplitude.}
The results in Table~\ref{tab:results} clearly show that this task is harder than \emph{Left-Right}. 
%The naive baseline here is given by predicting the mean angle and amplitude across the training set, and is $1.90$ radians for angle and $141.8$ pixels for the amplitude (RMSE in both cases).
Except for RBF SVC/SVR, which performs close to random, all classical machine learning models reach a similar performance for the estimation of both amplitude and angle. This result is above the naive baseline although not by a big margin. Tree-based models perform the best among classical statistical models, however, they are clearly inferior to deep learning models in this task. 
%Somewhat surprisingly, KNN is the best performing classical method with an RMSE of $1.48$ radians for the angle and $117.5$ pixels for the amplitude.
%However, the gap with ideal performance is still very large. In the case of angle estimation, the top scoring models are again tree-based, i.e., XGBoost, GradientBoosting and RandomForest, and the performance difference with respect to the other models studied is considerable. Surprisingly, the performance in amplitude estimation follows a relatively different trend, with simpler models like ridge and lasso regression outperforming XGBoost. 
%On the other hand, deep learning models are clearly superior in this task, especially in amplitude estimation where their RMSE is much smaller than the naive baseline. 
Somewhat surprisingly, among the deep learning models the simplest ones, i.e., CNN and PyramidalCNN perform the best, with an RMSE of $0.33$ and $0.34$ radians in angle estimation, and $32$ and $30.7$ mm in amplitude estimation, respectively. 
Overall, we see that there is still a considerable gap between the best results reported here and ideal performance. Our results constitute a baseline that should orient future work on estimating angle and amplitude of saccades from EEG data.
%All of the evaluated deep learning models obtain an RMSE in amplitude estimation close to $50$ pixels, except for EEGNet that obtains $65.1$. However, EEGNet obtains the best results in angle estimation, with an RMSE of $0.88$ radians. This suggests that estimating angle and amplitude may require different information, and thus, a model well-suited for angle estimation (e.g., EEGNet) does not necessarily perform accordingly when estimating the amplitude and viceversa (e.g. CNN). 
%Overall, we see that there is a considerable gap between the best results reported here and ideal performance. Our results constitute a baseline that should orient future work on estimating angle and amplitude of saccades from EEG data.

\paragraph{Absolute Position.}
Finally, we see in the last column of Table~\ref{tab:results} the results of our experimentation on absolute position estimation. We see that classical models generally fail in this task, with performances very close to the naive baseline.
On the other hand, deep learning models reach an euclidean distance with respect to the true location in the range of $70$ to $80$ mm. Although these results are far from ideal performance, the considerable gap with the naive baseline shows that EEG-based eye tracking can potentially be achieved to an acceptable degree of accuracy.
Our results, set the baseline for this task in $70.2$ mm, as reached by CNN, again one of the simpler deep learning models. However, as explained in Section~\ref{sec:absPos}, exploiting previous information is likely to improve performance, and thus, it would be interesting to see how sequence models, such as Recurrent Neural Networks or Transformers would perform in this task. There is a lot of room for improvement and we hope that this task will help future work in advancing EEG-based eye tracking.

In summary, our evaluation reveals that deep learning models are superior to other statistical techniques in estimating gaze position from EEG data. Although this is not surprising, given the complexity of the task and the larger expressive capacity of neural networks, it confirms that EEGEyeNet is a valuable resource for developing large neural models. We expect that future work will surpass our scores advancing EEG-based eye tracking.

%%%%%%%%%

\section{Discussion and Future Work}

The dataset and benchmark presented in this work provide an approachable framework to conduct research in the intersection of brain activity and eye movement. We will actively maintain and continuously extend both the dataset and the benchmark with further measurements from more participants and new experimental paradigms. We acknowledge as current limitations the relatively small number of participants recorded for the \emph{Large Grid} paradigm and the fact that the test sets of our benchmark are publicly available. We will address both points by (1) recording more data and (2) building an automatic evaluation service that keeps the test set hidden from the users and includes a leaderboard.

To facilitate extensive use of EEGEyeNet for various research purposes, we provide in our repository data preparation tools with an easy-to-use interface where users can define their own benchmarking tasks or extract other information from the dataset. 
In particular, the user can specify whether some recording blocks from the experimental paradigm should be ignored, which events to extract from the data and how the data should be preprocessed. Additionally, the user can also decide whether feature extraction should be performed or not. 
As with the other resources presented in this work, the data preparation module is in continuous development. We plan to adapt this software tool in the future according to the users' needs.
Overall, we expect that EEGEyeNet will become a central resource for a broad range of EEG and Eye-Tracking related research, specifically:

\paragraph{Research in Cognitive Area.}
EEGEyeNet's rich structure and high-density coverage of EEG and Eye-Tracking data may help advance other areas that study the combination of gaze position and brain activity to identify variations in attention, arousal and participant’s compliance with the task demands. Moreover, the behavioral information gained from eye tracking with the high temporal resolution and neurophysiological markers provided by EEG enables research of the perceptual, attentional, cognitive processes.

\paragraph{Benchmarking.} 
We expect the high quality, diversity and large scale of the EEGEyeNet dataset to be leveraged, in order to include new tasks in the proposed benchmark, as well as to build benchmarks for related domains. In particular, we plan to include segmentation tasks that evaluate the ability of a given model to detect and distinguish events such as fixations or saccades.

\section{Conclusion}
Recording eye-tracking data is a complex and expensive process that requires specialized hardware, trained operators and participants' consent. Collecting such data in combination with EEG adds an additional level of complexity to the data acquisition process. Therefore, behavioral research studying the combination of brain activity and eye movements is typically restricted by the lack of appropriate data. In this work, we have introduced EEGEyeNet, a large dataset of EEG and eye tracking data, with the view of making basic cognitive neuroscience research more approachable. Furthermore, given the potential benefits that EEG-based eye-tracking can bring in different domains, we have proposed a benchmark to facilitate the development of new methods tackling this challenge. Our experiments on this benchmark show that deep learning models perform better than classical statistical models. This confirms that the amount of data contained in EEGEyeNet is large enough to reliably train large neural models, which is a promising direction for further developing EEG-based eye tracking. 

\clearpage

\clearpage

\bibliographystyle{unsrtnat}
\bibliography{sample-base}
%\end{comment}
%%%%%%%%%%%%%%%%%%%%%%%%%%%%%%%%%%%%%%%%%%%%%%%%%%%%%%%%%%%%
%%%%%%%%%%%%%%%%%%%%%%%%%%%%%%%%%%%%%%%%%%%%%%%%%%%%%%%%%%%%

\clearpage

\appendix
\section*{Author statement}
Hereby we confirm that we bear all responsibility in case of violation of rights, etc., and confirmation of the data license.
All data are de-identified and participants gave permission for their data to be openly shared as part of the informed consent process.

Public data are distributed under the the Creative Commons Attribution 4.0 International Public License (\url{https://creativecommons.org/licenses/by/4.0/}).

We release our code used for our experiments, data collection and data preparation at the following repository : \url{https://github.com/ardkastrati/EEGEyeNet} and our dataset at  \url{https://doi.org/10.17605/OSF.IO/KTV7M}.

\section*{Acknowledgments}
We thank Tzvetan Popov for useful discussions related to this project
and Marta Marciniak for the  language editing, and proofreading. 
Dataset collection and maintenance is generously supported by the Velux Foundation and Swiss National Science Foundation.

We thank the participants for taking the time to be in the study, as well as their willingness to have their data shared with the scientific community.
\newpage

\section*{Supplementary Material 
}

\section{Preprocessing}\label{app:prep}
\paragraph{EEG Minimal Preprocessing} In this preprocessing strategy first, we identify bad channels using the EEGLAB \cite{delorme2004eeglab} clean rawdata algorithm \footnote{ \texttt{clean\_rawdata()} from: \url{http://sccn.ucsd.edu/wiki/Plugin_list_process}}. This algorithm consists of three steps: (1) it detects channels that have no signal variation for more than 5 seconds and filters the EEG of the remaining channels with a forward-backwards (non-causal) filter, here we use a high-pass filter with a transition band of [$0.25$, $0.75$] Hz; (2), it searches for channels with a lower correlation to its robust estimate than a given threshold that we set to $0.85$; and (3) it removes EEG channels with excessive line noise that in our case corresponds to $4$ standard deviations.
Subsequently, the high Variance Criterion (HVC) was applied, with a pre- specified   threshold of $100 \mu V$ as the upper limit for identifying bad channels.
Later,  the EEG data were band-pass filtered at $0.5$ and $40$ Hz and then, Zapline toolkit \cite{de2020zapline} was applied to remove line noise artifacts, discarding $7$ power line components. Finally, all the channels marked as bad were interpolated using spherical spline interpolation.

\paragraph{EEG Maximal Preprocessing.}
In addition to the whole minimal preprocessing pipeline, in the Maximal Preprocessing Pipeline, the Independent component analysis (ICA) is used to isolate the various source generator processes underlying those recordings. Non-Brain artifactual source components  are removed based on the automatic classification result as provided by Independent Component Label (ICLabel) \cite{pion2019iclabel}. 
ICA included Optimizing the ICA-based removal of ocular EEG artifacts from free viewing experiments (Dimigen 2020) as implemented in \citet{ped}.

\paragraph{Objective Quality Classification }
After the bad channel interpolation, all EEG datasets were rated with the objective quality criteria of the Automagic Toolbox \cite{ped}. Any data file rated as bad, meaning that the proportion of high-amplitude data points in the signal (> $30 \mu V$) was larger than 0.2, more than 20\% of the time points showed a variance larger than $15 \mu V$ across channels, 40\%  of the channels showed high variance ($15 \mu V$), or the ratio of bad channels was higher than 0.4 was not included in the further data processing.

%\paragraph{Electroncephalograpy and Eye-Tracking Synchronisation}
%After preprocessing (both minimally and maximally), the EEG and eye-tracking data were synchronized using “EYE EEG ” \cite{dimigen2011coregistration} to enable EEG analyses time-locked to the onsets of saccades, fixations, and other triggers, depending on the experimental paradigm. Synchronization quality was ensured by comparing the trigger latencies recorded in the EEG and eye-tracker data. All synchronization errors did not exceed 2 ms.

\section{Annotations}\label{app:ann}

Saccades are detected by the velocity and acceleration of the eye movements. We used the default system (SR Research, http://www.sr-research.com/) parameters to define saccades: an acceleration threshold of 8000° per second, a velocity threshold of 30° per second, and a deflection threshold of 0.1°.

Fixations were defined as time periods without saccades. 
Fixation might include small saccades (i.e., microsaccades), which fall below the threshold for saccade detection.
Furthermore, a blink was be regarded as a special case of a fixation, where the pupil diameter is either zero or outside a dynamically computed valid pupil, or the horizontal and vertical gaze positions are zero. 

 We extracted the following information about the saccades, fixations and blinks: start and end time, duration, coordinates of start positions on the computer screen in pixels, and for saccades additionally: end positions, and amplitudes.

\section{Events Characteristics}\label{app:sc}
In this appendix we show further details of each dataset. More precisely, for each experimental paradigm we present the distribution of the fixation duration, the distribution of the fixation positions, the distribution of the saccade amplitude (in pixels, where 1 pixel corresponds to 0.5 mm), and the distribution of the saccade angle. 
  
\begin{figure}[H]
\centering
\begin{subfigure}{.40\textwidth}
  \centering
  % include first image
  \includegraphics[width=\linewidth]{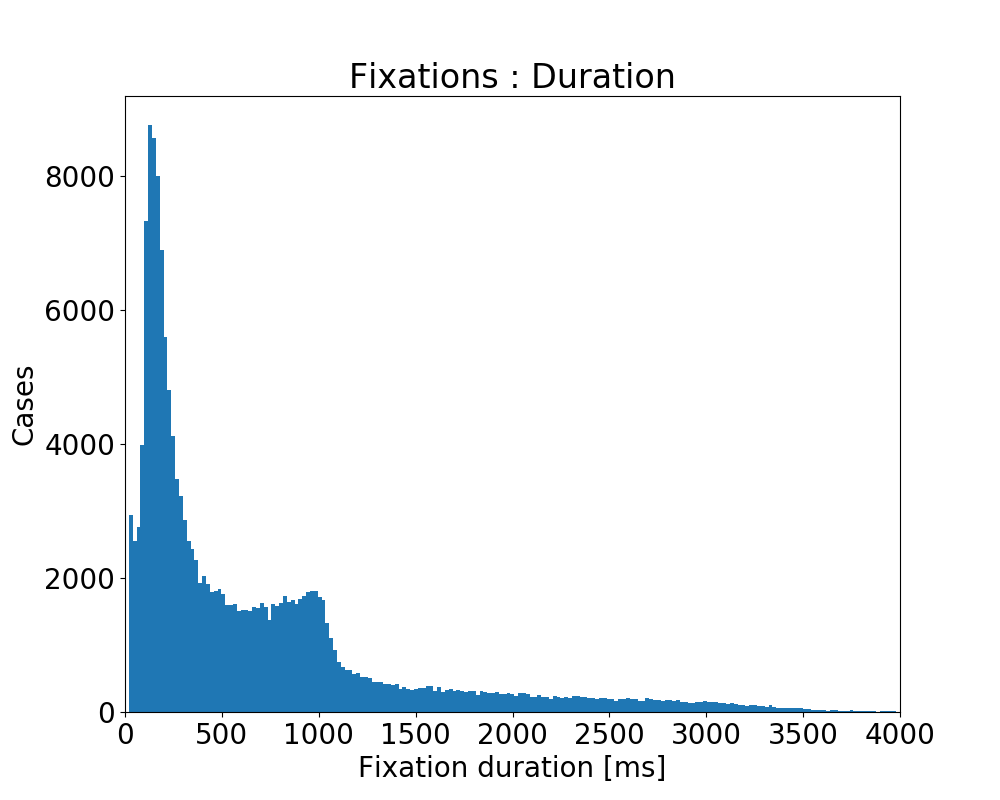}  
  \caption{Distribution of the fixation duration}
  \label{fig:sub-first}
\end{subfigure}
\hspace{2em}% Space between image A and B
\begin{subfigure}{.40\textwidth}
  \centering
  % include second image
  \includegraphics[width=\linewidth]{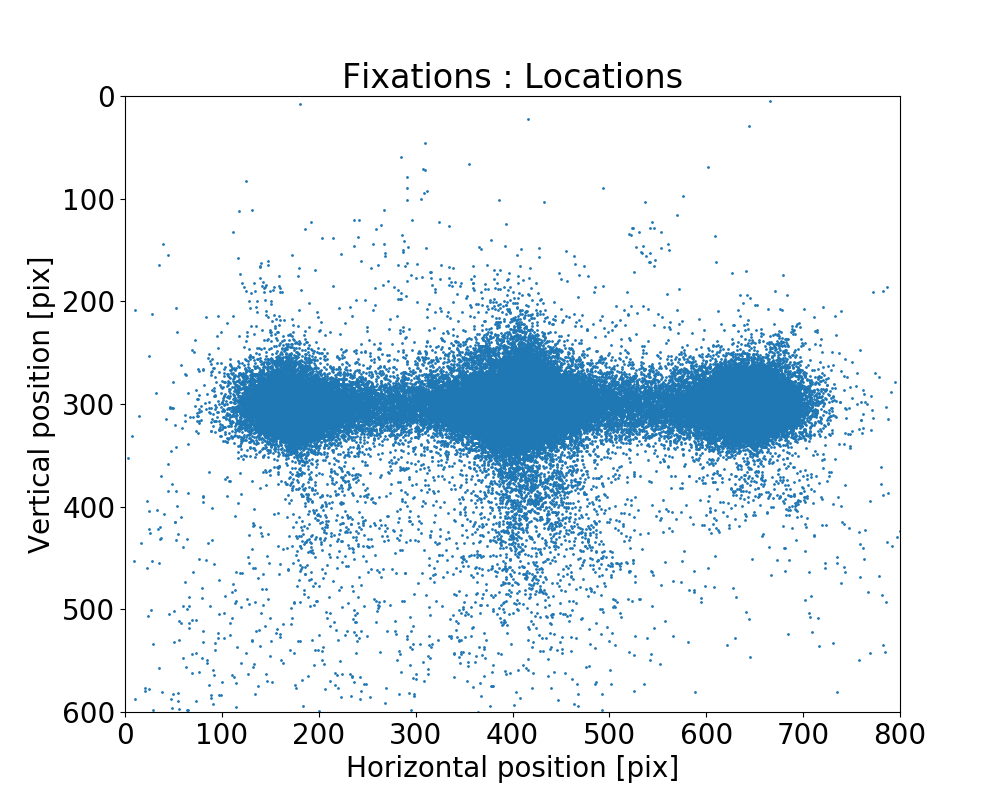}  
  \caption{Distribution of the fixation positions}
  \label{fig:sub-second}
\end{subfigure}

\begin{subfigure}{.40\textwidth}
  \centering
  % include third image
  \includegraphics[width=\linewidth]{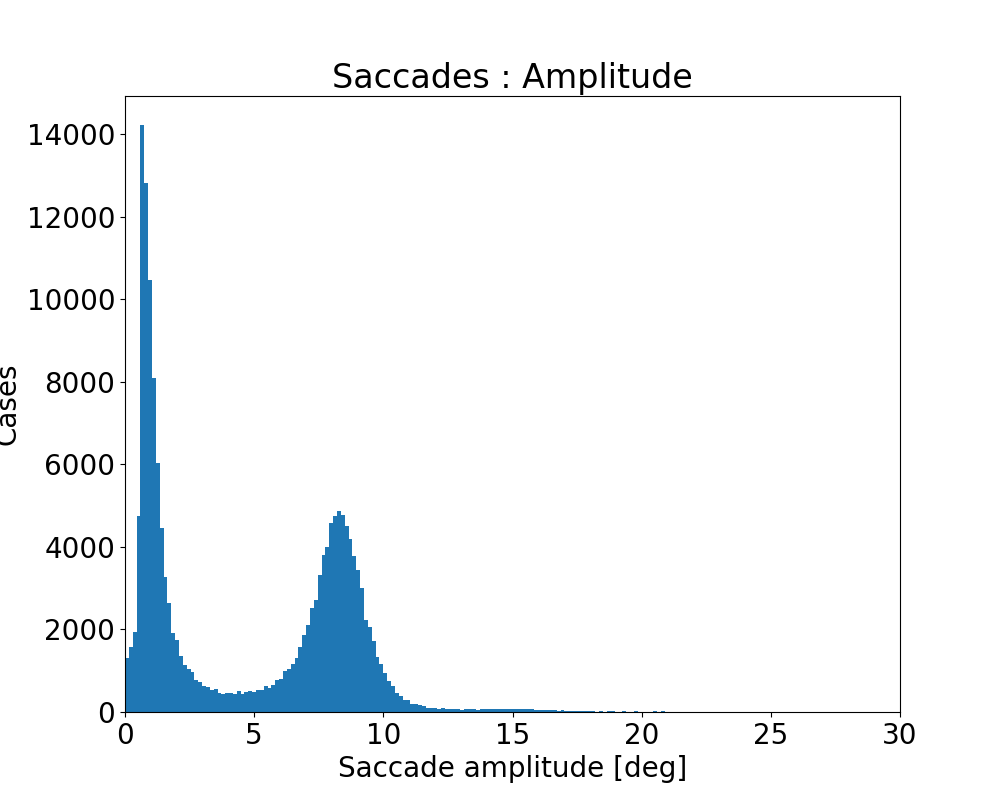}  
  \caption{Distribution of the saccade amplitude}
  \label{fig:sub-third}
\end{subfigure}
\hspace{2em}% Space between image A and B
\begin{subfigure}{.40\textwidth}
  \centering
  % include fourth image
  \includegraphics[width=\linewidth]{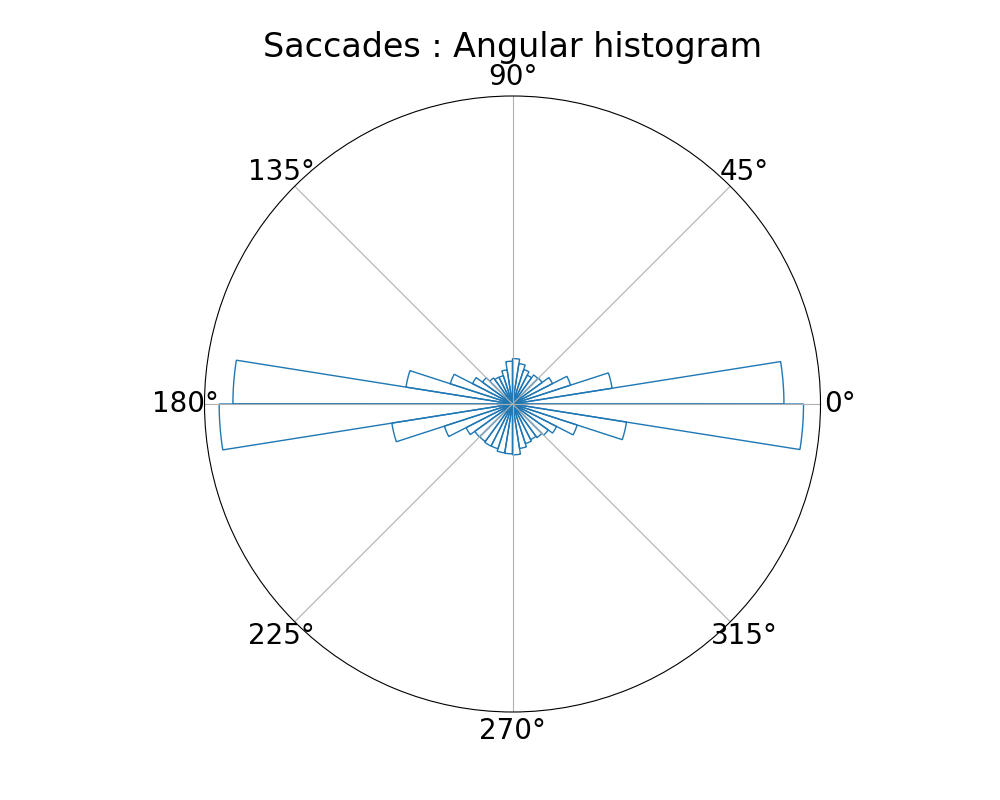}  
  \caption{Distribution of the saccade angle}
  \label{fig:sub-fourth}
\end{subfigure}
\caption{Antisaccade Paradigm}
\label{fig:antisaccade:distribution}
\end{figure}

\begin{figure}[H]
\centering
\begin{subfigure}{.40\textwidth}
  \centering
  % include first image
  \includegraphics[width=\linewidth]{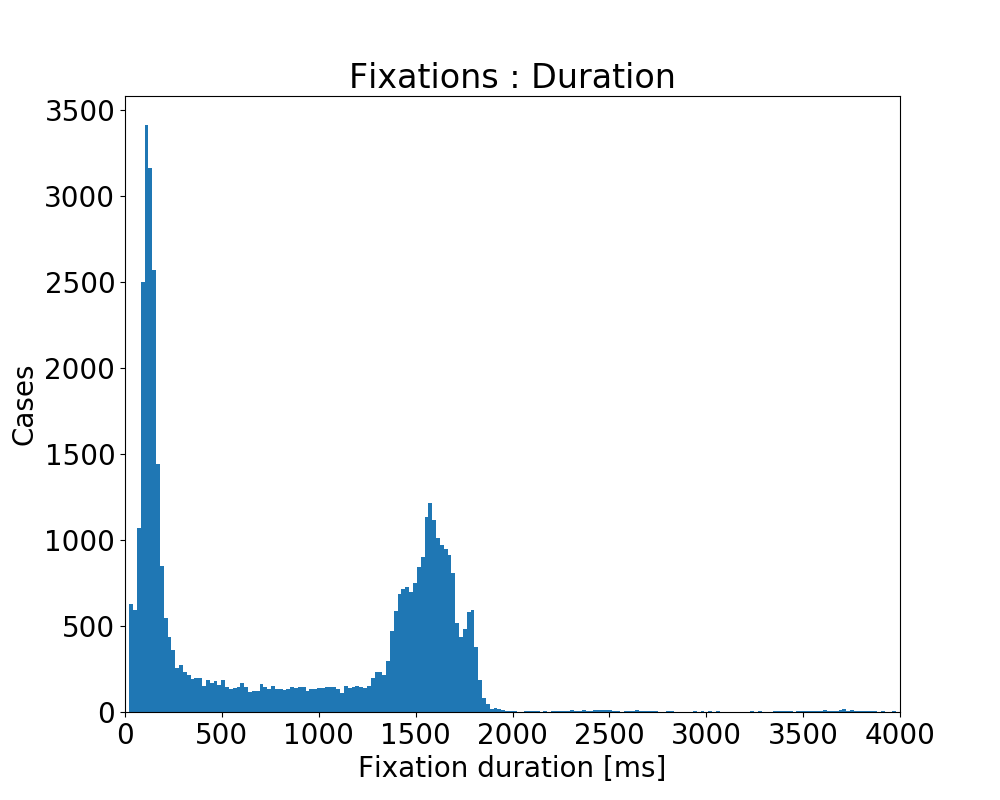}
 \caption{Distribution of the fixation duration}
  \label{fig:sub-first}
\end{subfigure}
\hspace{2em}% Space between image A and B
\begin{subfigure}{.40\textwidth}
  \centering
  % include second image
  \includegraphics[width=\linewidth]{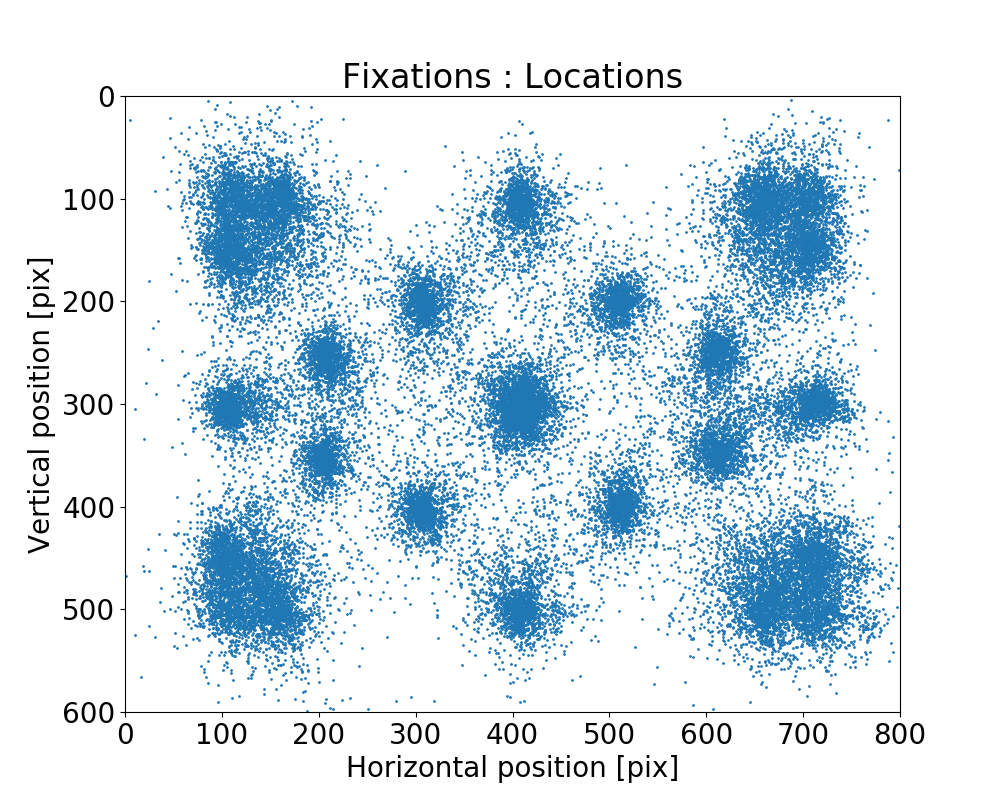}
  \caption{Distribution of the fixation positions}
  \label{fig:sub-second}
\end{subfigure}

\begin{subfigure}{.40\textwidth}
  \centering
  % include third image
  \includegraphics[width=\linewidth]{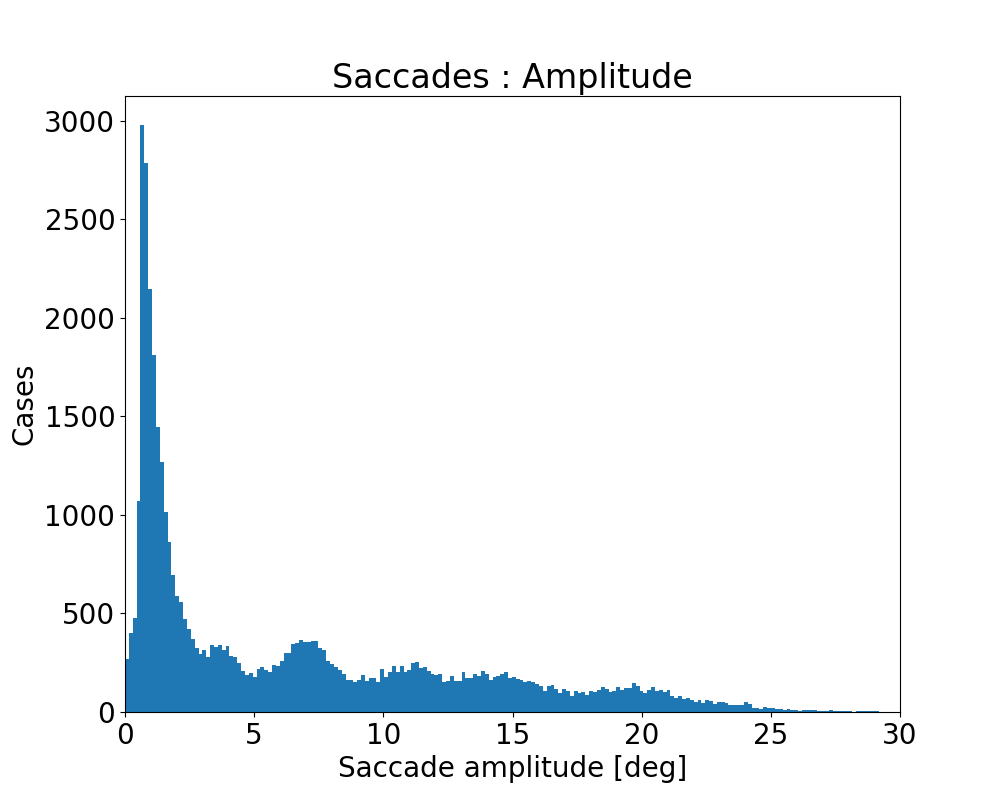}
 \caption{Distribution of the saccade amplitude}
  \label{fig:sub-third}
\end{subfigure}
\hspace{2em}% Space between image A and B
\begin{subfigure}{.40\textwidth}
  \centering
  \includegraphics[width=\linewidth]{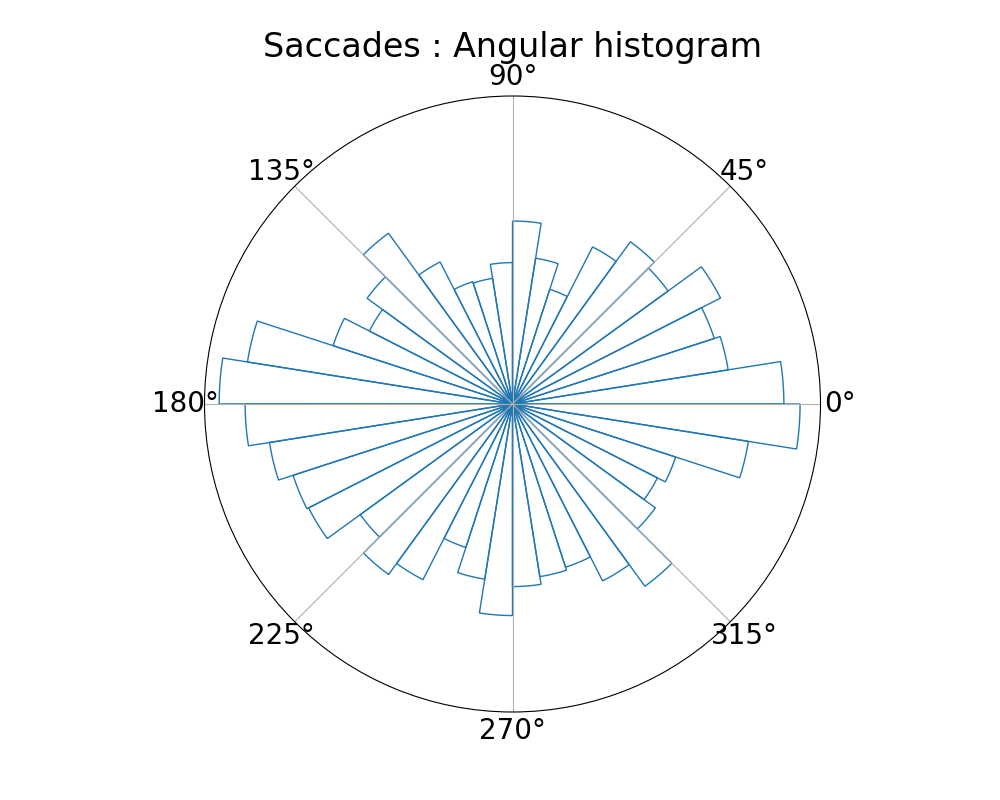}
  \caption{Distribution of the saccade angle}
  \label{fig:sub-fourth}
\end{subfigure}
\caption{Large Grid Paradigm}
\label{fig:large-grid:distribution}
\end{figure}

\clearpage
  \begin{figure}[H]
  \centering
\begin{subfigure}{.40\textwidth}
  \centering
  % include first image
  \includegraphics[width=\linewidth]{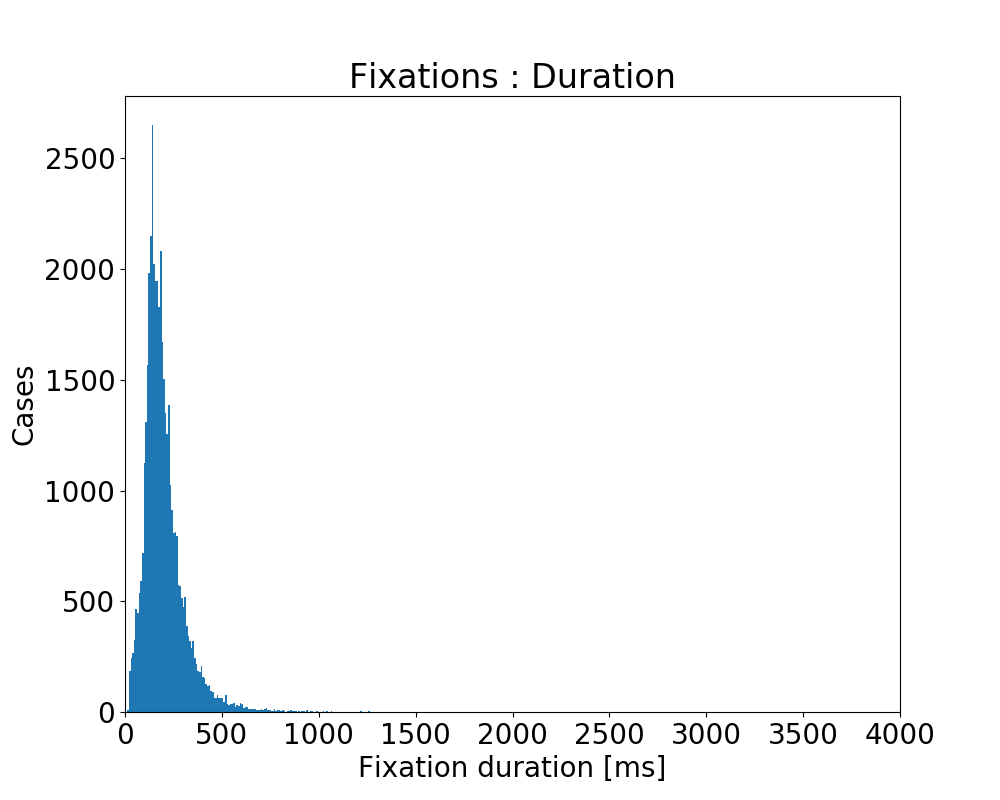}
  \caption{Distribution of the fixation duration}
  \label{fig:sub-first}
\end{subfigure}
\hspace{2em}% Space between image A and B
\begin{subfigure}{.40\textwidth}
  \centering
  % include second image
  \includegraphics[width=\linewidth]{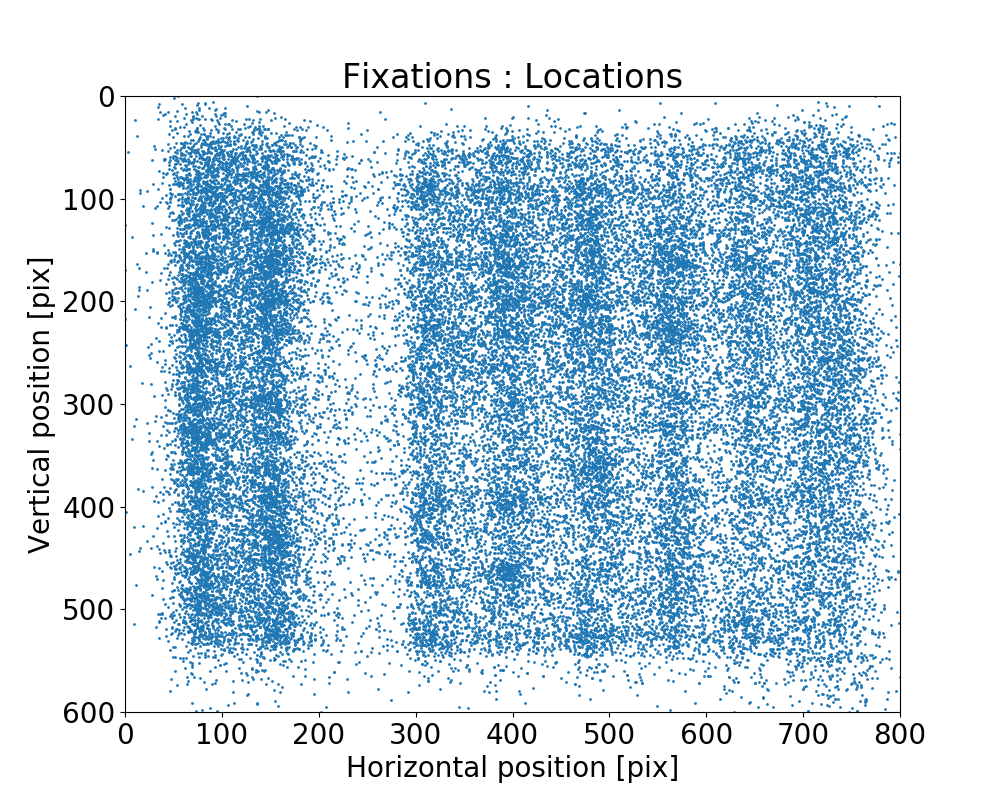}
 \caption{Distribution of the fixation positions}
  \label{fig:sub-second}
\end{subfigure}

\begin{subfigure}{.40\textwidth}
  \centering
  % include third image
  \includegraphics[width=\linewidth]{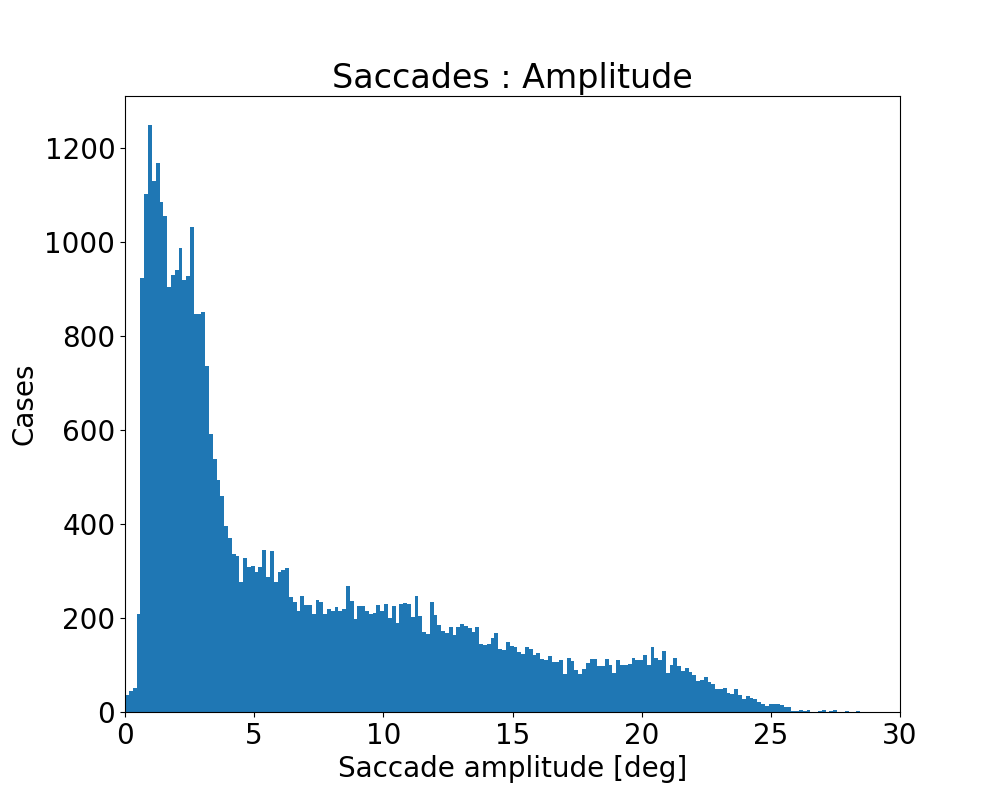}
  \caption{Distribution of the saccade amplitude}
  \label{fig:sub-third}
\end{subfigure}
\hspace{2em}% Space between image A and B
\begin{subfigure}{.40\textwidth}
  \centering
  % include fourth image
  \includegraphics[width=\linewidth]{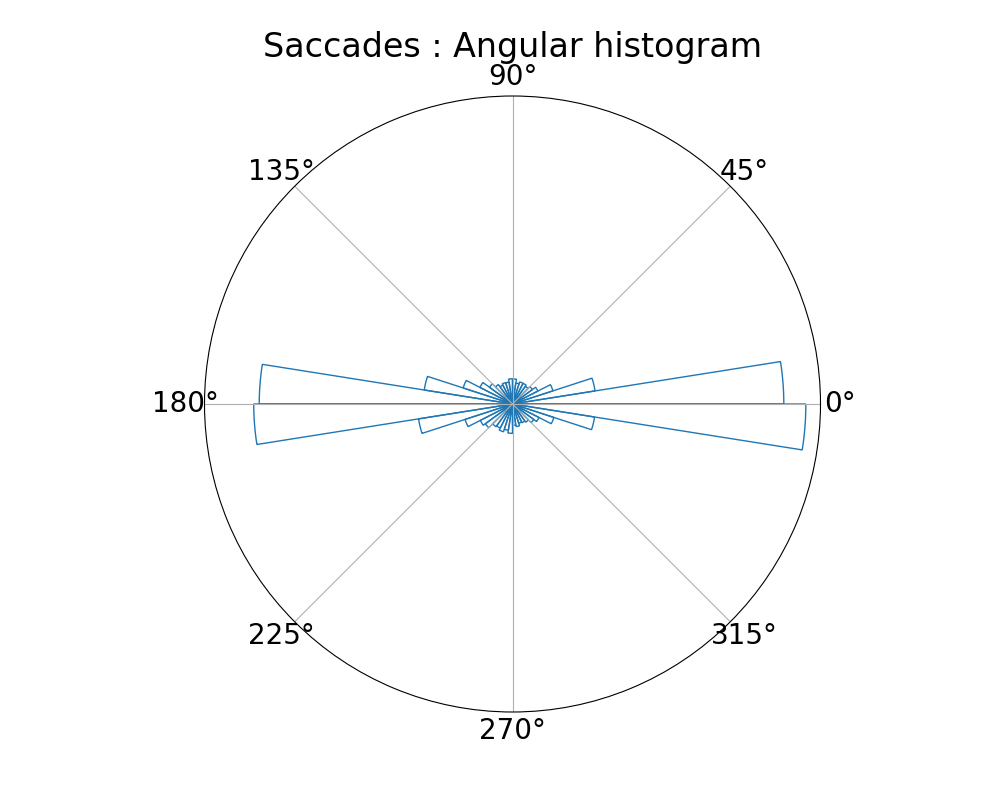}
  \caption{Distribution of the saccade angle}
  \label{fig:sub-fourth}
\end{subfigure}
\caption{Visual Symbol Search Paradigm}
\label{fig:vss:distribution}
\end{figure}

 \begin{figure}[H]
  \centering
\begin{subfigure}{.40\textwidth}
  \centering
  % include first image
  \includegraphics[width=\linewidth]{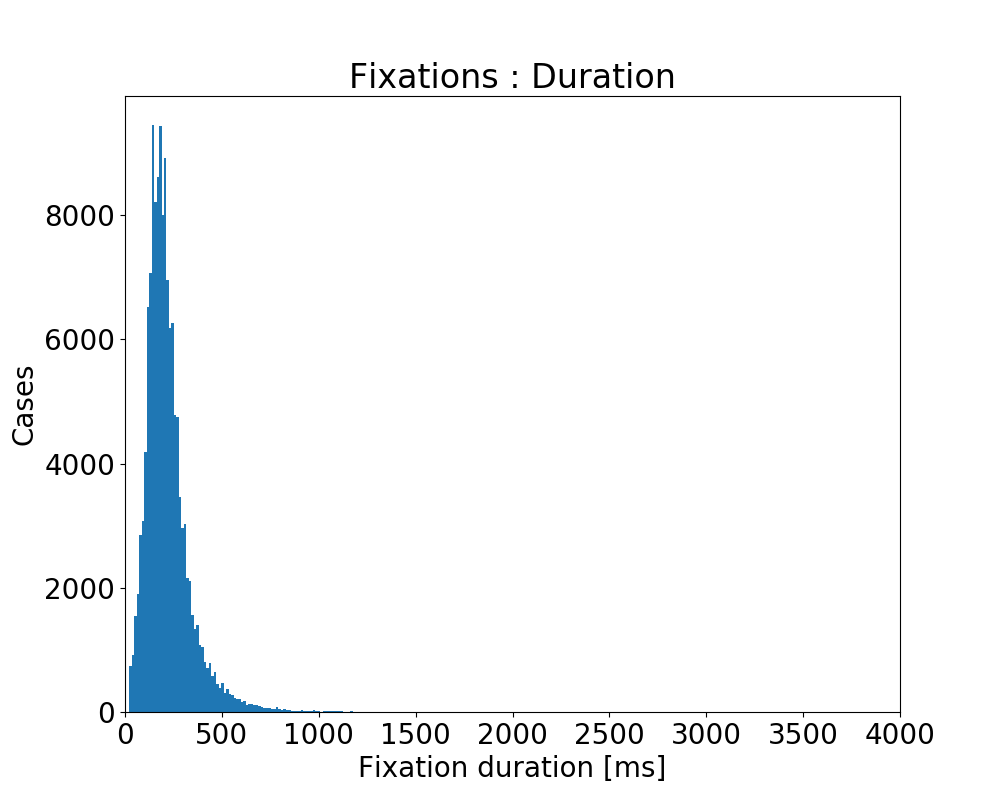}
  \caption{Distribution of the fixation duration}
  \label{fig:sub-first}
\end{subfigure}
\hspace{2em}% Space between image A and B
\begin{subfigure}{.40\textwidth}
  \centering
  % include second image
  \includegraphics[width=\linewidth]{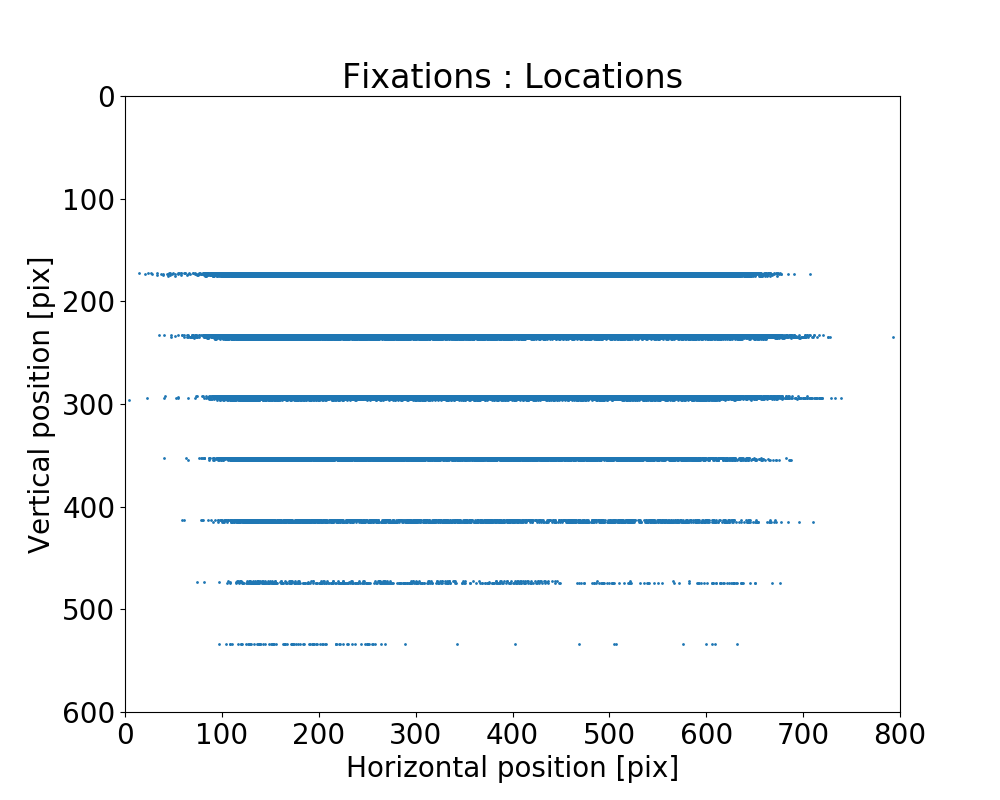}
 \caption{Distribution of the fixation positions}
  \label{fig:sub-second}
\end{subfigure}

\begin{subfigure}{.40\textwidth}
  \centering
  % include third image
  \includegraphics[width=\linewidth]{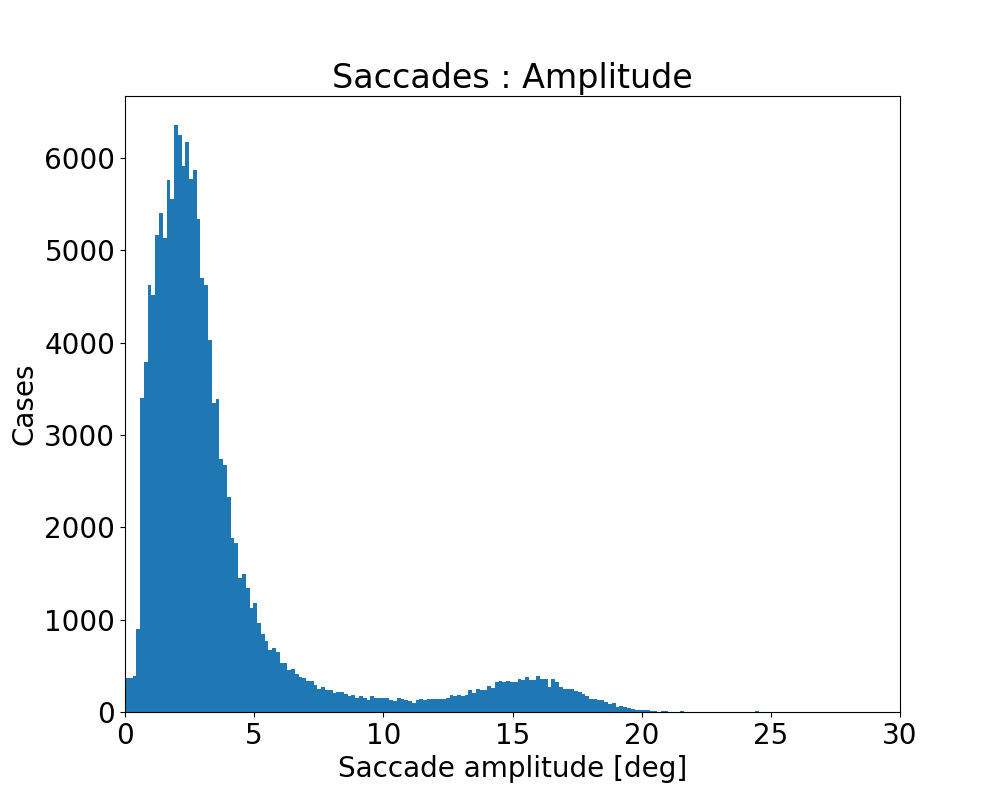}
  \caption{Distribution of the saccade amplitude}
  \label{fig:sub-third}
\end{subfigure}
\hspace{2em}% Space between image A and B
\begin{subfigure}{.40\textwidth}
  \centering
  % include fourth image
  \includegraphics[width=\linewidth]{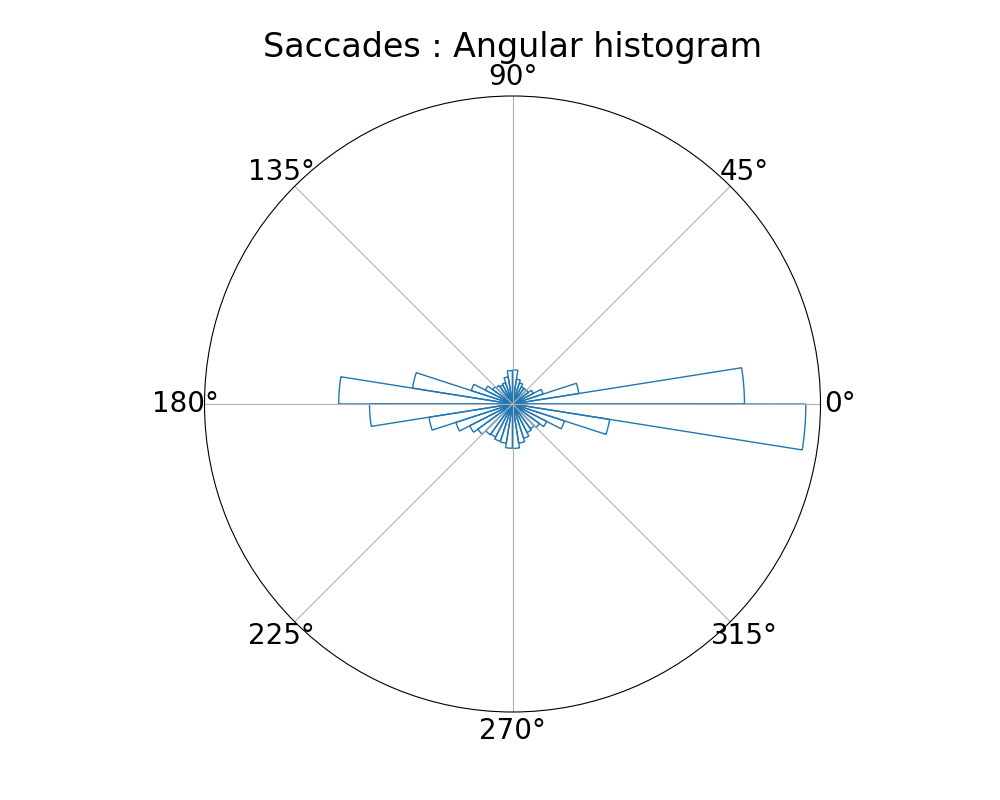}
  \caption{Distribution of the saccade angle}
  \label{fig:sub-fourth}
\end{subfigure}
\caption{ZuCo 2.0 Paradigm}
\label{fig:zuco:distribution}
\end{figure}

 \section{Details of the Models}\label{app:DL}

In this appendix we present the details of our models used during our experimentations.

\subsection{Deep Learning Models}
In the following we present five different deep learning architectures that we evaluate on the EEGEyeNet benchmark.

\paragraph{EEGNet.} \cite{eegnet} is a convolutional neural network tailored to Brain-Computer Interfaces applications. This model first performs a temporal convolution to learn frequency filters, then a depthwise convolution to learn frequency-specific spatial filters followed by a separable convolution which should learn a temporal summary for each individual feature map. Finally, a pointwise convolution is applied to mix the feature maps. For details on this architecture we refer the reader to the original work \cite{eegnet}, which we follow in our implementation.

\paragraph{Convolutional Neural Network (CNN).} We experiment with a standard one-dimensional convolutional neural network with $12$ layers and additive residual connections around blocks of three layers. Each layer consists of (1D-)convolution, batch normalization \cite{ioffe2015batch}, ReLU activation and max pooling. In the convolutions we use $16$ filters of size size $64$, and for the pooling operation a kernel of size $2$ and stride $1$. Each residual connection performs a convolution followed by batch normalization. 

\paragraph{Pyramidal CNN.} We consider a CNN with an inverted pyramidal shape and $6$ layers. Each layer consists of the same modules as the CNN, with the difference that the number of filters is a multiple of $16$ that grows with depth, i.e., $16$ in the first layer, $32$ in the second and so on. In this case, the kernel size is $16$ and there are not residual connections.

\paragraph{InceptionTime} We re-implement the model from~\cite{fawaz2020inceptiontime}, which is an adaptation of the Inception-v4 architecture \cite{szegedy2017inception} to time series classification. We instantiate this model with $12$ layers and skip connections every three layers. Each layer consists of an InceptionTime module, which takes as input $64$ channels and performs a $1\times 1$ bottleneck convolution that produces a feature map with $16$ channels. Then, this feature map is passed through three different convolution operators with $16$ filters each and kernel sizes of $64$, $32$, $16$; and a max-pooling operator with a kernel size of $3$. The four resulting feature maps are concatenated to form the layer's output, which has $64$ channels.This model also uses residual connections around each block of three layers.

\paragraph{Xception} Finally, we build a model based on the Xception architecture proposed by \citet{Chollet_2017_CVPR}. This model follows the same general structure as our CNN, with $12$ layers and residual connections after three layers. Each layer contains a 1D depthwise separable convolution~\cite{Chollet_2017_CVPR}, with $64$ filters and kernel size of $40$, followed by batch normalization and ReLU activation.

\begin{table*}[h]
\centering
\scalebox{0.7}{
\begin{tabular}{@{}l*{7}{S[table-format=-3.4]}@{}}
\toprule
{Model} & {$Par 1$} & {$Par 2$} & {$Par 3$} & {$Par 4$}  \\
\midrule  
{KNN} & \makebox{leaf size 10} & \makebox{10LR/25AN/50AP} & \makebox{-} & \makebox{-}  \\
{GaussianNB} & \makebox{var smooth: 0.0004941713} & \makebox{-} & \makebox{-} & \makebox{-} \\
{LinearSVC} & \makebox{C:0.01} & \makebox{tol: 1e-05} & \makebox{max iter: 1200} & \makebox{-}  \\
{RBF SVC/SVR} & \makebox{C:1} & \makebox{tol: 1e-05} & \makebox{max iter: 1200} & \makebox{gam:0.01}  \\
{Linear regression} & \makebox{all default} & \makebox{-} & \makebox{-} & \makebox{-}  \\
{Lasso regression} & \makebox{alpha: 0.01AN/1} & \makebox{tol: 1e-05} & \makebox{max iter: 1200} & \makebox{-}  \\
{Elastic Net} & \makebox{alpha: 0.1AN/1} & \makebox{l1 ratio: 0.9LR/0.3AN/0.6AP,} & \makebox{tol: 1e-05} & \makebox{gam:0.01}  \\
{Random Forest} & \makebox{max depth 10LR/10AN/50} & \makebox{est: 50AN/50AM/250} & \makebox{-} & \makebox{-}  \\
{AdaBoost} & \makebox{lr: 0.5LR/0.5AN/0.1AM/0.01AP} & \makebox{est: 100LR/50} & \makebox{-} & \makebox{-} \\
{XGBoost} & \makebox{eta: 0.05AN/0.1} & \makebox{est: 100AP/250} & \makebox{max depth: 10AP/5}  & \makebox{-} & 
\\
%& \makebox{0 \rpm 0.0} & \makebox{0 \rpm 0.0} & \makebox{0 \rpm 0.0} & \makebox{0 \rpm 0} \\   
\bottomrule
\end{tabular}
}
\caption{Hyperparameters of classical machine learning models. LR - Left/Right task, AN - Angle task, AM - Amplitude task, and AP - Absolute Position task.}  
\label{tab:hyperparameters:classical}
\end{table*}

\begin{table*}[h]
\centering
\scalebox{0.95}{
\begin{tabular}{@{}l*{7}{S[table-format=-3.4]}@{}}
\toprule
{Parameter} & {$CNN$} & {$PyramidalCNN$} & {$EEGNet$} & {$InceptionTime$} & {$Xception$}  \\
\midrule  
{Depth} & \makebox{12} & \makebox{6} & \makebox{2} & \makebox{12} & \makebox{18} \\
{Number of filters} & \makebox{16} & \makebox{16} & \makebox{16/256} & \makebox{16} & \makebox{64} \\
{Kernel size} & \makebox{64} & \makebox{16} & \makebox{256} & \makebox{64} & \makebox{40}\\
{Batch size} & \makebox{64} & \makebox{64} & \makebox{64} & \makebox{64} & \makebox{64} \\
{Epochs} & \makebox{50} & \makebox{50} & \makebox{50} & \makebox{50} & \makebox{50} \\
{Early stopping patience} & \makebox{20} & \makebox{20} & \makebox{20} & \makebox{20} & \makebox{20} \\
{Residual connections} & \makebox{True} & \makebox{False} & \makebox{False} & \makebox{True} & \makebox{True} \\
{Bottleneck size} & \makebox{-} & \makebox{-} &  \makebox{-} & \makebox{16} &  \makebox{-} \\
{Dropout rate} &  \makebox{-} &  \makebox{-} & \makebox{0.5} &  \makebox{-} &  \makebox{-} \\
%& \makebox{0 \rpm 0.0} & \makebox{0 \rpm 0.0} & \makebox{0 \rpm 0.0} & \makebox{0 \rpm 0} \\   
\bottomrule
\end{tabular}
}
\caption{Hyperparameters of the deep learning models. EEGNet uses two different amounts of filters in its two blocks.}  
\label{tab:hyperparameters:deep-learning}
\end{table*}

\subsection{Hyperparameter Tuning}

In Table \ref{tab:hyperparameters:classical} we report our tuned hyperparameters for the classical machine learning models and in Table \ref{tab:hyperparameters:deep-learning} for the deep learning models.

\section{Further Experiments}\label{app:FurthExp}

\subsection{Experiments on the Maximally Preprocessed Data}
Besides the minimally preprocessed data presented in the benchmark results at Section \ref{sec:benchmark}, we performed the same experiments on the maximally preprocessed data. As expected, the models perform much worse compared to the ones with minimally preprocessed data (see Table \ref{tab:results:maximal}). However, the following results show that even after removing eye artifacts during preprocessing, it is still possible to infer gaze direction based on EEG data.

\begin{table*}[h]
\centering
\begin{tabular}{@{}lc*{5}{S[table-format=-3.4]}@{}}
\toprule
\toprule
& {Left-Right} & \multicolumn{2}{c}{Angle/Amplitude} & {Abs. Position}\\
\cmidrule(lr){2-2} \cmidrule(lr){3-4} \cmidrule(lr){5-5}
{Model} & {Accuracy} & {Angle RMSE} & {Amp. RMSE} & {RMSE} \\
\midrule  
{KNN} & \makebox{60.9 \rpm 0} & \makebox{1.83 \rpm 0} & \makebox{72.2 \rpm 0} & \makebox{123.1 \rpm 0}\\
{GaussianNB} & \makebox{58.5 \rpm 0} & \makebox{-} & \makebox{-} & \makebox{-}\\
{LinearSVC} & \makebox{63.2 \rpm 0} & \makebox{-} & \makebox{-} & \makebox{-}\\
{RBF SVC/SVR} & \makebox{53.9 \rpm 0} & \makebox{1.91 \rpm 0} & \makebox{76.1 \rpm 0} & \makebox{122.6 \rpm 0}\\
{Linear Regression} & \makebox{-} & \makebox{1.76 \rpm 0} & \makebox{72.2 \rpm 0} & \makebox{123.5 \rpm 0}\\
{Ridge Regression} & \makebox{-} & \makebox{1.79 \rpm 0} & \makebox{71.6 \rpm 0} & \makebox{122.6 \rpm 0}\\
{Lasso Regression} & \makebox{-} & \makebox{1.77 \rpm 0} & \makebox{71.4 \rpm 0} & \makebox{122.6 \rpm 0}\\
{Elastic Net} & \makebox{-} & \makebox{1.80 \rpm 0} & \makebox{71.5 \rpm 0} & \makebox{122.5 \rpm 0}\\
\midrule
%{Decision Tree} & \makebox{96.2 \rpm 0} & \makebox\\
{Random Forest} & \makebox{66.1 \rpm 0.2} & \makebox{1.77 \rpm 0.01} & \makebox{71.1 \rpm 0} &  \makebox{123.5 \rpm 0.2} \\
{Gradient Boost} & \makebox{66.9 \rpm 0.2} & \makebox{1.78 \rpm 0.01} & \makebox{71 \rpm 0} &  \makebox{122.4 \rpm 0}\\
{AdaBoost} & \makebox{65.9 \rpm 0} & \makebox{1.84 \rpm 0.01} & \makebox{72.8 \rpm 0} & \makebox{122.7 \rpm 0.1}\\
{XGBoost} & \makebox{68.3 \rpm 0} & \makebox{1.75 \rpm 0} & \makebox{70.5 \rpm 0} & \makebox{123.2 \rpm 0} \\
\midrule
{CNN} & \makebox{75.5 \rpm 1.7} & \makebox{\textbf{1.07} \rpm 0.05} & \makebox{76.8 \rpm 2.3} & \makebox{134.6 \rpm 2.8}\\
{PyramidalCNN} & \makebox{\textbf{86.6} \rpm 1.1}  & \makebox{1.38 \rpm 0.35} & \makebox{76.6 \rpm 2.2} & \makebox{142.1 \rpm 3.1} \\
{EEGNet} & \makebox{83.6 \rpm 1.1}  & \makebox{1.31 \rpm 0.03} & \makebox{\textbf{67.8}\rpm 2.1} & \makebox{\textbf{119.9} \rpm 0.8}\\
{InceptionTime} & \makebox{80.4 \rpm 2.9} & \makebox{1.65 \rpm 0.33} & \makebox{70.8 \rpm 1.7} & \makebox{137.2 \rpm 1.7}\\
{Xception} & \makebox{75.7 \rpm 1.8} & \makebox{1.57 \rpm 0.29} & \makebox{76.3 \rpm 2.1} & \makebox{141.1 \rpm 1.3}\\
\midrule
{Naive Baseline} & \makebox{52.3} & \makebox{1.90} & \makebox{74.7} & \makebox{123.3} \\
\bottomrule
\end{tabular}
\caption{Data maximally preprocessed, 5 runs per DL model, Adam optimizer with learning rate 1e-4, early stopping patience 20. \emph{Angle} is measured in radians, \emph{Amplitude} and \emph{Abs. Position} in mm.
}  
\label{tab:results:maximal}
\end{table*}

\paragraph{Left-Right.} 
We see that classical machine learning models achieve a better performance than the naive baseline of $52.26\%$. In particular, tree-based models (RandomForest, GradientBoosting, AdaBoost and XGBoost) reach a performance of over $68\%$. Although it is not a high performance, it shows that the eye movement information between left and right can also be inferred from ``pure'' brain activity. Furthermore, the deep learning models achieve an even higher performance of over $86\%$, showing that one can achieve satisfactory results for this task also from the maximally preprocessed data.
% reason for the high performance in this task, is the large amount of data that we make available in our dataset, which allows the models to fully exploit their expressive capacity, unlike in related datasets, where the best performing models reached an accuracy of only $X\%$. 
% Sanity check?

\paragraph{Angle/Amplitude.}
As in the case for minimally preprocessed data (see Table \ref{tab:results}), the results in Table~\ref{tab:results:maximal} show that this task is more demanding than Left-Right. All classical models perform close to random. The deep learning models perform slightly better, and, interestingly, the simple CNN architecture performs best. This result is above the naive baseline but not by a big margin. Nonetheless, it shows that information about the saccade angle can be extracted from maximally preprocessed data as well.
In comparison to the angle, we see that the amplitude task is strictly more difficult for maximally preprocessed data in the third column. All models except EEGNet fail at this task, with EEGNet performing slightly better than the naive baseline. This leaves the open question to what extent the angle and amplitude of saccades can be estimated from maximally preprocessed EEG data. %, which shows that there is room for improvements in this task as well.

\paragraph{Absolute Position.}
In the last column of Table~\ref{tab:results:maximal} we see that all models fail in this task, with performances very close to the naive baseline.  Deep learning models (except EEGNet) perform even worse than the naive baseline. At this point, it is not clear whether the absolute position of fixations can be inferred from the maximally preprocessed EEG data.

\subsection{Experiments on ZuCo 2.0 Dataset}

For comparison, we performed the same experiments on one of these openly available datasets, namely the ZuCo 2.0. We have used the same pipeline from our infrastructure. First, we have synchronized the EEG-ET data and extracted the relevant events for annotation (fixation and saccades). Then we also performed feature extraction. Afterwards, we used our data preparation tool to extract the samples for our benchmarking tasks. Finally, we ran our benchmark with this dataset. Note that we didn't use the LR-task for this dataset due to the nature of the ZuCo 2.0 experimental paradigm. Although during reading, most of the eye movements are towards the right, the dataset contains eye movements in all other directions (see Figure \ref{fig:zuco:distribution}). 
Other referenced datasets don't provide an interface for benchmarking purposes. Again, this is because there are differences in the recording setup, protocol and paradigms.

\begin{table*}[h]
\centering
\begin{tabular}{@{}lc*{5}{S[table-format=-3.4]}@{}}
\toprule
\toprule
& \multicolumn{2}{c}{Angle/Amplitude} & {Abs. Position}\\
\cmidrule(lr){2-2} \cmidrule(lr){3-4} \cmidrule(lr){5-5}
{Model} & {Angle RMSE} & {Amp. RMSE} & {RMSE} 
\\
\midrule  
{KNN}  & \makebox{1.42 \rpm 0} & \makebox{59.9\rpm 0} & \makebox{72.2 \rpm 0}
\\
%{GaussianNB}  & \makebox{-} & \makebox{-} & \makebox{-}\\
%{LinearSVC}  & \makebox{-} & \makebox{-} & \makebox{-}\\
{RBF SVC/SVR}  & \makebox{1.47 \rpm 0} & \makebox{91.4 \rpm 0} & \makebox{77.5 \rpm 0}
\\
{Linear Regression}  & \makebox{1.40 \rpm 0} & \makebox{67.6 \rpm 0} & \makebox{72.2 \rpm 0}
\\
{Ridge Regression}  & \makebox{1.41 \rpm 0} & \makebox{67.4 \rpm 0} & \makebox{72 \rpm 0}
\\
{Lasso Regression}  & \makebox{1.41 \rpm 0} & \makebox{67.3 \rpm 0} & \makebox{72 \rpm 0}
\\
{Elastic Net}  & \makebox{1.42 \rpm 0} & \makebox{67.3 \rpm 0} & \makebox{72 \rpm 0}
\\
\midrule
%{Decision Tree} & \makebox{96.2 \rpm 0} & \makebox\\
{Random Forest}  & \makebox{1.42 \rpm 0.01} & \makebox{70.6 \rpm 0} &  \makebox{72.9 \rpm 0.1} 
\\
{Gradient Boost}  & \makebox{1.42 \rpm 0.01} & \makebox{67.4 \rpm 0} &  \makebox{61.9 \rpm 0}
\\
{AdaBoost}  & \makebox{1.41 \rpm 0.01} & \makebox{69 \rpm 0} & \makebox{75.3 \rpm 0.2}
\\
{XGBoost}  & \makebox{1.42 \rpm 0} & \makebox{67.5 \rpm 0} & \makebox{\textbf{72.1} \rpm 0} \\
\midrule
\midrule  
{CNN} & \makebox{1.54 \rpm 0.4} & \makebox{67.1 \rpm 1.1} & \makebox{75.8 \rpm 0.3} 
\\
{PyramidalCNN}  & \makebox{1.70 \rpm 0.4} & \makebox{\textbf{61.9} \rpm 0.6} & \makebox{73.8 \rpm 1.6} 
\\
{EEGNet}  & \makebox{\textbf{1.35} \rpm 0.2} & \makebox{62.1 \rpm 1.5} & \makebox{72.7 \rpm 0.1} 
\\
{InceptionTime} & \makebox{1.81 \rpm 0.02} & \makebox{67.3 \rpm 1.1} & \makebox{80 \rpm 1.1} 
\\
{Xception}  & \makebox{1.67 \rpm 0.4} & \makebox{65.9 \rpm 1.9} & \makebox{76.2 \rpm 2.4} 
\\
\midrule
{Naive Baseline} & \makebox{1.41} & \makebox{67} & \makebox{72} \\
\bottomrule
\end{tabular}
\caption{Results of Angle/Amplitude and Absolute Position tasks on the ZuCo dataset. Same hyperparameters as for experiments on our dataset. \emph{Angle} is measured in radians, \emph{Amplitude} and \emph{Abs. Position} in mm.
} 
\label{tab:results:zuco}
\end{table*}  

As expected, the models perform worse compared to the ones with minimally preprocessed data (see Table \ref{tab:results}).

\paragraph{Angle/Amplitude.}
We can see a clear difference between the results in our dataset (see Table \ref{tab:results}) and the results obtained in ZuCo 2.0 dataset. The models performs worse in the ZuCo 2.0 dataset. For the angle task, the best model is the EEGNet with RMSE 1.35 radians. It performs better then the naive baseline (RMSE 1.90 radians), but not by a big margin. In comparison, in our dataset, EEGNet performs much better with an RMSE of 0.70 radians. In addition, the best performing model for our dataset is the simple CNN, which has an RMSE of 0.33 radians. We can also see a clear difference for the amplitude task as well. The best performing model for ZuCo 2.0 dataset is PyramidalCNN with an RMSE of $61.9$ mm, while in our dataset PyramidalCNN has a better performance with an RMSE of $30.7$ mm.

\paragraph{Absolute Position.}
In this task, the results are not directly comparable to our dataset, since they depend on the distribution of the fixation positions. The difference of the distributions can be seen in Figure \ref{fig:large-grid:distribution} and Figure \ref{fig:zuco:distribution}. Furthermore, in Table \ref{tab:results} we can see that the naive baseline for our dataset is 123.3, whereas for the ZuCo 2.0 dataset (Table \ref{tab:results:zuco}), it is 72. In other words, in ZuCo 2.0 simply predicting the mean position leads to a better performance compared to our dataset, because the fixations are not scattered across all possible positions on the screen. In contrast, our dataset is specifically tailored for eye movement prediction. However, we can still see a clear difference in the performance. In our dataset, there is a considerable gap between the deep learning models and the naive baseline, whereas in the ZuCo dataset all models perform close to the baseline. One possible explanation for this (besides the size and the type of dataset) is that ZuCo dataset uses a different preprocessing method, whereas our dataset uses minimally preprocessing data  which is more suitable for eye movement prediction.

\section{Datasheets for Datasets}\label{app:datasheet}
For the dataset documentation we used the recommended
documentation framework "Datasheets for Datasets" \cite{gebru2018datasheets}.

\subsection{Motivation} \paragraph{For what purpose was the dataset created?}
The primary purpose of the project is to advance research that studies the combination of brain activities and gaze position.
\paragraph{Who created the dataset} The dataset was collected by the Methods of Plasticity Research Lab at the University of Zurich. All persons participating in the data collection are acknowledged
on our website: www.eegeye.net
\paragraph{Who funded the creation of the dataset? If there is an associated
grant, please provide the name of the grantor and the grant name and
number}
This work was supported by the Velux Stiftung Project No. 1126 and by the Schweizerischer Nationalfonds zur Förderung der Wissenschaftlichen Forschung (SNF) Grant $100014_175875$.
\paragraph{Any other comments?} \answerNA{}

\subsection{Composition}
\paragraph{What do the instances that comprise the dataset represent (e.g.,
documents, photos, people, countries)? Are there multiple types of
instances (e.g., movies, users, and ratings; people and interactions between them; nodes and edges)?}
the EEGEyeNet dataset.
 contains electroencephalography and eye-tracking  recordings following three different experimental paradigms. 
Together with the raw data, we release two sets of preprocessed data: minimally and maximally preprocessed; as well as the preprocessing code.
\paragraph{How many instances are there in total (of each type, if appropriate)?}
The number of instances is presented in the Table \ref{tab:benStats} (Benchmark statistic).
\paragraph{Does the dataset contain all possible instances or is it a sample
(not necessarily random) of instances from a larger set?} The dataset used in this project was recorded in our laboratory in the context of a larger project that aims to quantify age effects on eye movement behaviour and electroencephalography (EEG) recordings of resting-state and task-related activity. Therefore, for the EEGEyeNet dataset, we have used three paradigms that can help in the gaze position prediction.

%Inclusion criteria for participation in the study were left and right-handedness, healthy male and female participants, with age between 20 and 80 years old. Before each recording, we assured that the participant did not suffer from psychiatric symptoms, had no severe neurologic disorders, prior head injuries, a stroke, a transient circulatory disorder of the brain, diagnosis of dementia  Huntington’s disease, Parkinson’s disease, sensory and/or motor problems that interfere with computer tasks. Moreover, exclusion criteria were current use of psychotropic drugs, intake of recreational synthetic or natural drugs.
\paragraph{What data does each instance consist of? “Raw” data (e.g., unprocessed text or images) or features?} For each experimental paradigm, as described in the section \ref{sec:para}, we release raw data, two sets of preprocessed data: minimally and maximally preprocessed; as well as the preprocessing code. 
\paragraph{Is there a label or target associated with each instance? If so, please
provide a description}\answerYes{All labels and targets are described in the Chapter \ref{sec:benchmark} }.
\paragraph{Is any information missing from individual instances? If so, please
provide a description, explaining why this information is missing (e.g.,
because it was unavailable). This does not include intentionally removed
information, but might include, e.g., redacted text.}\answerNA{}
\paragraph{Are relationships between individual instances made explicit
(e.g., users’ movie ratings, social network links)? }  \answerNA{}
\paragraph{Are there recommended data splits (e.g., training, development/validation,
testing)?} \answerYes{All recommended splits with rationales behind them are reported in the chapter~\ref{sec:benchmark}}.
\paragraph{Are there any errors, sources of noise, or redundancies in the
dataset?} 
Raw data is often contaminated by artifacts influenced by technical and environmental factors and the recorded participant's specific. 

\paragraph{ Is the dataset self-contained, or does it link to or otherwise rely on
external resources (e.g., websites, tweets, other datasets)?} The dataset is self contained, there are no restrictions associated with any of the external resources that
might apply to a future user.
\paragraph{Does the dataset contain data that might be considered confidential }\answerNo{The dataset was anonymized}.
\paragraph{Does the dataset contain data that, if viewed directly, might be offensive, insulting, threatening, or might otherwise cause anxiety?} \answerNA{}.
\paragraph{Does the dataset relate to people?}\answerYes{}
\paragraph{Does the dataset identify any subpopulations (e.g., by age, gender)?} 
\answerYes{Subpopulations are described in the Subsection \ref{sec:dataset2}}.

%If so, please describe how these subpopulations are identified and
%provide a description of their respective distributions within the dataset.
\paragraph{Is it possible to identify individuals (i.e., one or more natural persons), either directly or indirectly (i.e., in combination with other
data) from the dataset?}  \answerNo{}.
\paragraph{Does the dataset contain data that might be considered sensitive
in any way }
%(e.g., data that reveals racial or ethnic origins, sexual
%orientations, religious beliefs, political opinions or union memberships, or locations; financial or health data; biometric or genetic data; forms of government identification, such as social security numbers; criminal history)? 
\answerNo{}
\paragraph{Any other comments?} \answerNA{}

\subsection{Collection Process}
\paragraph{How was the data associated with each instance acquired? Was
the data directly observable (e.g., raw text, movie ratings), reported by
subjects (e.g., survey responses), or indirectly inferred/derived from other
data (e.g., part-of-speech tags, model-based guesses for age or language)?}
We recorded ET and EEG brain activity, all procedures and mechanims are described in the Section \ref{sec:dataset}.
\paragraph{What mechanisms or procedures were used to collect the data
(e.g., hardware apparatus or sensor, manual human curation, software program, software API)? How were these mechanisms or procedures validated?}
All procedures and mechanims are described in the Section \ref{sec:dataset}.
\paragraph{If the dataset is a sample from a larger set, what was the sampling
strategy (e.g., deterministic, probabilistic with specific sampling
probabilities)?}\answerNA{}
\paragraph{Who was involved in the data collection process (e.g., students,
crowdworkers, contractors) and how were they compensated (e.g.,
how much were crowdworkers paid)?}
Research Assistants and PhD Students were involved in the data collection process. All obtained fixed salary for Research Assistants/PhD Students.
\paragraph{Over what timeframe was the data collected?} The collection of data used in the project began in the first quarter of 2018 and lasted until May 15, 2021.
\paragraph{Were any ethical review processes conducted (e.g., by an institutional review board)?} 
This study was conducted according to the principles expressed in the Declaration of Helsinki. The study was approved by the Institutional Review Board of Canton Zurich (BASEC-Nr. 2017-00226).
\paragraph{Does the dataset relate to people?} \answerYes{}
\paragraph{ Did you collect the data from the individuals in question directly,
or obtain it via third parties or other sources (e.g., websites)?} We collected the data from the individuals in question directly. 
\paragraph{Were the individuals in question notified about the data collection?} 
\answerYes{All participants gave their written informed consent before participation in the study, as provided in the Appendix "Formal consent of Participants".}
\paragraph{Did the individuals in question consent to the collection and use
of their data?} \answerYes{
The template of the formal consent can be found in the Appendix \ref{app:ee}.}
\paragraph{If consent was obtained, were the consenting individuals provided with a mechanism to revoke their consent in the future or
for certain uses?}\answerYes{The Participants signed the following agreement:"I decide voluntarily and can withdraw this decision at any time. If I no longer want to participate, my data will be irrevocably deleted.
I only inform the project management and do not have to justify this decision".
The German version of the agreement can be found in the Appendix \ref{app:ee}.}
\paragraph{Has an analysis of the potential impact of the dataset and its use
on data subjects (e.g., a data protection impact analysis)been conducted?}\answerNo{}
\paragraph{Any other comments?} \answerYes{
 Prior to visiting the laboratory, participants completed a 10 min. pre-screening interview over the phone with a research assistant to confirm their eligibility and safety to participate in the study. This brief interview obtains information regarding an individual’s psychiatric history, including past or present diagnoses and/or treatment, as well as current medications and any neurological disorders (see \ref{app:inc}). If a participant demonstrates no contraindications for EEG (e.g., history of seizures or epilepsy), he or she is then scheduled for a research study appointment.}

\subsection{Preprocessing/cleaning/labeling}
\paragraph{Was any preprocessing/cleaning/labeling of the data done (e.g.,
discretization or bucketing, tokenization, part-of-speech tagging,
SIFT feature extraction, removal of instances, processing of missing values)?}\answerYes{All preprocessing steps are described in the Dataset section} 
remainder of the questions in this section.
\paragraph{Was the “raw” data saved in addition to the preprocessed/cleaned/labeled
data (e.g., to support unanticipated future uses)}
\answerYes{Together with the raw data, we release two sets of preprocessed data: minimally and maximally preprocessed; as well as the preprocessing code. 
This way, we give a user the freedom to manipulate raw data while easing the experimentation barrier by additionally providing ready-to-use clean data.}
\paragraph{Is the software used to preprocess/clean/label the instances available? If so, please provide a link or other access point.}\answerYes{}
\paragraph{Any other comments?} \answerNA{}
\subsection{Uses}
\paragraph{Has the dataset been used for any tasks already?}\answerYes{A subset of the Antisaccade Paradigm was already used in the publication \cite{martynka}}. 
\paragraph{Is there a repository that links to any or all papers or systems that
use the dataset?} \answerNo{This repository will be created and maintained in the future.}
\paragraph{What (other) tasks could the dataset be used for?}
The dataset includes  experimental paradigms assessing key cognitive functions, like inhibitory control and processing speed. Additionally, the dataset can be used for a development of new segmentation algorithms for fixations, saccades, and blinks.
\paragraph{Is there anything about the composition of the dataset or the way
it was collected and preprocessed/cleaned/labeled that might impact future uses?}
\answerYes{As described in the preprocessing section of the manuscript, depending on how the data is processed and whether the eye movements component is retained, we may receive different results.}
%%%
\paragraph{Are there tasks for which the dataset should not be used?} \answerNo{}
\paragraph{Any other comments?} \answerNA{}
\subsection{Distribution}
\paragraph{ Will the dataset be distributed to third parties outside of the entity (e.g., company, institution, organization) on behalf of which
the dataset was created?} \answerNo{}
\paragraph{How will the dataset will be distributed (e.g., tarball on website,
API, GitHub)? Does the dataset have a digital object identifier (DOI)?}
%%%
The dataset will be made available along with the manuscript on our website.
Raw and preprocessed EEG and eye-tracking data are available online and have the digital object identifier: DOI 10.17605/OSF.IO/KTV7M
\paragraph{When will the dataset be distributed?}
The dataset will be distributed along with the manuscript submission.
\paragraph{Will the dataset be distributed under a copyright or other intellectual property (IP) license, and/or under applicable terms of use
(ToU)?} 
The dataset will distributed under the terms of the Creative Commons CC BY license, which permits unrestricted use, distribution, and reproduction in any medium, provided the original work is properly cited.
\paragraph{Have any third parties imposed IP-based or other restrictions on
the data associated with the instances?} \answerNo{}
\paragraph{Do any export controls or other regulatory restrictions apply to
the dataset or to individual instances?}\answerNo{}
\paragraph{Any other comments} \answerNA{}
\subsection{Maintenance}
\paragraph{Who is supporting/hosting/maintaining the dataset?}
The dataset will be maintanted by the EEGEyeNet team.
\paragraph{How can the owner/curator/manager of the dataset be contacted
(e.g., email address)?}
Email addres: akastrati@ethz.ch, martyna.plomecka@uzh.ch
\paragraph{Is there an erratum? If so, please provide a link or other access point} \answerNo{}
\paragraph{Will the dataset be updated (e.g., to correct labeling errors, add
new instances, delete instances)?} 
%\
\answerYes{The dataset will be regularly updated, information about each update will be published in our repository, available on the website www.eegeye.net.}
\paragraph{If the dataset relates to people, are there applicable limits on the
retention of the data associated with the instances (e.g., were individuals in question told that their data would be retained for a
fixed period of time and then deleted)?} \answerNo{}.
\paragraph{Will older versions of the dataset continue to be supported/hosted/maintained?
}
\answerYes{Yes, all published versions of the dataset will be on our webpage www.eegeye.net.}
\paragraph{If others want to extend/augment/build on/contribute to the
dataset, is there a mechanism for them to do so? }
\answerYes{We release our complete infrastructure and provide a simple and easy-to-use interface to evaluate new methods.}
\paragraph{Any other comments?}
 \answerNA{}
    
\section{Risk categorisation}\label{app:risk}
Our study was classified as "Risk category A" by the Institutional Review Board of Canton Zurich (BASEC-Nr. 2017-00226). Applied methods of EEG and eye-tracking are non-invasive and do not pose risk to participants. The involved tasks do not involve deception and completion of these do not pose any harm to participants.

\section{Inclusion and Exclusion criteria}\label{app:inc}
Inclusion criteria for participation in the study were left and right-handedness, healthy male and female participants, with age between 20 and 80 years old. Before each recording, we assured that the participant did not suffer from psychiatric symptoms, had no severe neurologic disorders, prior head injuries, a stroke, a transient circulatory disorder of the brain, diagnosis of dementia  Huntington’s disease, Parkinson’s disease, sensory and/or motor problems that interfere with computer tasks. Moreover, exclusion criteria were current use of psychotropic drugs, intake of recreational synthetic or natural drugs.

\section{Potential negative societal impacts of the work}\label{app:neg}
In the specific case of our dataset, one risk of EEG-based eye-tracking models is that they could allow the tracking of the gaze of a subject without their consent, i.e., if the subject agrees to get their EEG data analyzed but not their gaze patterns. Furthermore, a major ethical concern is data privacy. A large-scale dataset such as the one we release in this work could be used in the future to develop methods capable of inferring sensitive information from EEG recordings, such as medical conditions. For this reason, we do not release any privacy-sensitive data.
Especially, the anonymity of the participants will be guaranteed when presenting the data at scientific meetings or publishing them in scientific journals. Interested third parties can access the data (but under no circumstances the personal data) via the research repository, solely for scientific purposes (e.g. replication or further analysis).
Individual participant medical information obtained from this research project is considered confidential, and disclosure to third parties is prohibited. Participant confidentiality will be further ensured by utilizing identification code numbers to correspond to medical information in the computer files.
    
\section{Formal consent of Participants}\label{app:ee}
In the following pages we present the template of the formal consent given to each participant. Every participant signed the formal consent before taking part in this study.


\section*{Supplementary material}

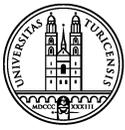
Psychologisches Institut
Methoden der Plastizitätsforschung

**Einwilligungserklärung**

**Schriftliche Einwilligungserklärung zur Teilnahme an einem Studienprojekt**
Bitte lesen Sie dieses Formular sorgfältig durch. Bitte fragen Sie den Versuchsleiter/die Versuchsleiterin, wenn Sie etwas nicht verstehen oder wissen möchten.

| | |
|---|---|
| **BASEC-Nummer (nach Einreichung):** | 2017-00226 |
| **Titel der Studie (wissenschaftlich und Laiensprache):** | Prediction of eye-motion with EEG |
| **verantwortliche Institution (Projektleitung mit Adresse):** | Psychologisches Institut<br>Universität Zürich<br>Andreasstrasse 15<br>8050 Zürich |
| **Ort der Durchführung:** | Binzmühlstrasse 14<br>8050 Zürich |
| **Leiter / Leiterin der Studie am Studienort:**<br>Name und Vorname in Druckbuchstaben: | |
| **Teilnehmerin/Teilnehmer:**<br>Name und Vorname in Druckbuchstaben:<br>Geburtsdatum: | ☐ weiblich    ☐ männlich |

- Ich wurde über den Zweck, den Ablauf des Projekts, über mögliche Vor- und Nachteile sowie über eventuelle Risiken informiert.
- Ich nehme an diesem Projekt freiwillig teil und akzeptiere den Inhalt der zum oben genannten Projekt abgegebenen schriftlichen Information. Ich hatte genügend Zeit, meine Entscheidung zu treffen.
- Meine Fragen im Zusammenhang mit der Teilnahme an diesem Projekt sind mir beantwortet worden. Ich behalte die schriftliche Information und erhalte eine Kopie meiner schriftlichen Einwilligungserklärung.
- Ich weiss, dass meine Daten in verschlüsselter Form zu Forschungszwecken weitergegeben werden können (auch ins Ausland).
- Ich kann jederzeit und ohne Angabe von Gründen von der Teilnahme zurücktreten, ohne dass ich deswegen Nachteile habe.
- Ich bin mir bewusst, dass die in der Teilnehmerinformation genannten Pflichten einzuhalten sind.

| Ort, Datum | Unterschrift Teilnehmerin/Teilnehmer |
|---|---|
| | |

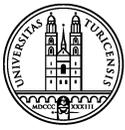
**Psychologisches Institut**
Methoden der Plastizitätsforschung

**Bestätigung der Prüfperson:** Hiermit bestätige ich, dass ich dieser Teilnehmerin/ diesem Teilnehmer Wesen, Bedeutung und Tragweite des Projekts erläutert habe. Ich versichere, alle im Zusammenhang mit diesem Projekt stehenden Verpflichtungen gemäss des geltenden Rechts zu erfüllen. Sollte ich zu irgendeinem Zeitpunkt während der Durchführung des Projekts von Aspekten erfahren, welche die Bereitschaft der Teilnehmerin/ des Teilnehmers zur Teilnahme an der Studie beeinflussen könnten, werde ich sie/ ihn umgehend darüber informieren.

| Ort, Datum | Name und Vorname der informierenden der informierenden Prüfperson in Druckbuchstaben |
|---|---|
|  | Unterschrift der Prüfperson |
|  |  |

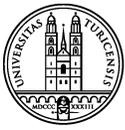

**Einwilligungserklärung für Weiterverwendung von Daten in verschlüsselter Form**

**Teilnehmerin/Teilnehmer:**
Name und Vorname in Druckbuchstaben:

Geburtsdatum: ☐ weiblich ☐ männlich

Ich erlaube, dass meine Daten in verschlüsselter Form für die Forschung (nicht kommerziell) weiter verwendet werden dürfen. Dies bedeutet, dass die Daten in verschlüsselter Form auf einem Server gespeichert sind und für zukünftige, noch nicht näher definierte Forschungsprojekte auf unbestimmte Zeitdauer verwendet werden dürfen.

Ich entscheide freiwillig und kann diesen Entscheid zu jedem Zeitpunkt wieder zurücknehmen. Wenn ich nicht mehr mitmachen möchte, werden meine Daten unwiderruflich gelöscht. Ich informiere lediglich die Projektleitung und muss diesen Entscheid nicht begründen.

Ich habe verstanden, dass die Daten verschlüsselt sind und der Schüssel sicher aufbewahrt wird. Die Daten können im In- und Ausland an andere (nicht kommerzielle) Forschungsgruppen zur Analyse gesendet werden, wenn diese dieselben Standards wie in der Schweiz einhalten. Alle rechtlichen Vorgaben zum Datenschutz werden eingehalten.

Normalerweise werden alle Daten gesamthaft ausgewertet und die Ergebnisse zusammenfassend publiziert. Die Ergebnisse aus den Daten werden nicht kommerziell genutzt werden.

| Ort, Datum | Unterschrift Teilnehmerin/ Teilnehmer |
|---|---|
| | |

**Bestätigung des Prüfperson:** Hiermit bestätige ich, dass ich dieser Teilnehmerin/ diesem Teilnehmer Wesen, Bedeutung und Tragweite der Weiterverwendung von Daten erläutert habe.

| Ort, Datum | Name und Vorname der der informierenden Prüfperson in Druckbuchstaben |
|---|---|
| | Unterschrift der Prüfperson |



%old, not included
%Furthermore, data recorded from participants of the study was excluded from the analysis if one of the following criteria were met: ET calibration failure, i.e., more than one visual degrees deviation on average across nine random visual stimulus presentations, EEG data evaluated as bad quality dataset (described in the Appendix 2), more than 50 \% of the data rejected during the data preparation phase (for each task the percentage was calculated separately, specific exclusion criteria are defined in the data preparation section).

\end{document}